\documentclass[iop,numberedappendix]{emulateapj}

\newcommand{\Revised}[1]{#1}

\usepackage{comment}
\usepackage{ulem}
\usepackage{tikz}
\usepackage{color}
\usepackage{graphicx}
\usepackage{amsmath}
\usepackage{amsfonts}

\newcommand{\beq}{\begin{equation}}
\newcommand{\eeq}{\end{equation}}
\newcommand{\bea}{\begin{eqnarray}}
\newcommand{\eea}{\end{eqnarray}}

\newcommand{\gv}[1]{\ensuremath{\mbox{\boldmath$ #1 $}}} 
\newcommand{\trm}[1]{\textrm{#1}}

\newcommand{\grad}[1]{\gv{\nabla} #1} 

\newcommand{\T}{\ensuremath{\tau_{E}}}
\newcommand{\aveT}{\ensuremath{\langle\tau_{E}\rangle}}
\newcommand{\Q}{\ensuremath{Q_\ast^\prime}}
\newcommand{\Porb}{\ensuremath{P}}
\newcommand{\Mj}{\ensuremath{M_{\rm J}}}
\newcommand{\Mp}{\ensuremath{M_p}}

\DeclareGraphicsExtensions{.png,.pdf}

\begin{document}

\shortauthors{essick and weinberg}
\shorttitle{orbital decay of hot jupiters}

\title{
Orbital decay of hot jupiters due to nonlinear tidal dissipation\\ within solar-type hosts
}

\author{Reed Essick and Nevin N.~Weinberg}
\affil{Department of Physics, and Kavli Institute for Astrophysics and Space Research, Massachusetts Institute of Technology,\\ Cambridge, MA 02139, USA}


\begin{abstract}
	We study the orbital evolution of hot Jupiters due to the excitation and damping of tidally driven $g$-modes within solar-type host stars.  
	Linearly resonant $g$-modes (the dynamical tide) are driven to such large amplitudes in the stellar core that they excite a sea of other $g$-modes through weakly nonlinear interactions.    
	By solving the dynamics of large networks of nonlinearly coupled modes, we show that the nonlinear dissipation rate of the dynamical tide is several orders of magnitude larger than the linear dissipation rate.     
	\Revised{We find  stellar tidal quality factors  $\Q \simeq 10^5-10^6$ for systems with planet} mass $\Mp \ga 0.5\Mj$ and \Revised{orbital} period $\Porb \la 2\trm{ days}$, \Revised{which implies that such systems} decay on timescales that are small compared to the main-sequence lifetime of their solar-type hosts.  
	\Revised{According to our results,} there are $\simeq10$ currently known exoplanetary systems,  including WASP-19b and HAT-P-36-b, with orbital decay timescales shorter than a Gyr. 
	Rapid, \Revised{tidally} induced orbital decay may explain the observed paucity of planets with $\Mp \ga \Mj$ and $P<2\trm{ days}$ around solar-type hosts and could generate detectable transit-timing variations in the near future.
\end{abstract}

\section{Introduction}\label{s:introduction}

The tide raised by a hot Jupiter excites large amplitude waves within its host star.  
These waves transfer energy and angular momentum from the orbit to the star and as a result the planet gradually spirals inward. 
The rate of orbital decay is determined by the efficiency of tidal dissipation and depends on the amplitude of the waves as well as the effectiveness of frictional processes within the star.  

Tidal dissipation is often parameterized by the stellar tidal quality factor \Q, where larger \Q\ implies less dissipation.  
Perhaps the best constraints on $\Q$ for solar-type stars come from the observed circularization rate of solar-type binaries, which yield $\Q \sim 10^6$ \citep{Meibom:2005}.  
However, because $\Q$ is not a fundamental property of the star (it depends on the shape and size of the orbit and the mass of the perturber), this result does not necessarily imply  $\Q \sim 10^6$ for hot Jupiter systems. 
There have been a number of efforts to measure $\Q$ from statistical modeling of the observed sample of hot Jupiters (see \citealt{Ogilvie:2014} for a review).  
\Revised{\citet{Penev:2012}} find that the distribution favors $\Q  \ga 10^7$ for a specific set of assumptions about the initial conditions. 
\citet{Jackson:2008} find \Revised{a best fit at $\Q\sim 10^{5.5}$ although they do not rule out much larger values} and note that it is difficult to obtain \Revised{tight} constraints because of the limited sample size and uncertainties in the initial period distribution and stellar age.  
Although there are no direct observational measurements of \Q\ from individual hot Jupiter systems (e.g., from the detection of orbital decay), \citet{Jackson:2009} argue that the distribution shows evidence for ongoing removal and destruction by tides.  
In addition, \citet{Teitler:2014} propose that the observed dearth of close-in planets around fast-rotating stars \citep{McQuillan:2013} can be attributed to tidal ingestion of giant planets.

\Revised{Linear tidal driving by the planet resonantly excites short wavelength waves within the host star.  In solar-type stars, these ``primary" waves are excited near the radiative-convective interface since in this region their 
wavelengths become large and they can couple to the long length scale tidal potential.   
Although the primary waves initially have relatively small amplitudes and are thus well-described by linear theory, as they propagate towards the stellar center their amplitudes increase due to geometric focusing (i.e., in 
order to conserve WKB flux within an ever decreasing volume).}
In hot Jupiter systems, the primary \Revised{waves reach large amplitudes as they approach the stellar core and become nonlinear, exciting} many secondary waves through nonlinear wave-wave interactions (\citealt{Barker:2010, Barker:2011a}; \citealt{Weinberg:2012}, hereafter WAQB).  
These secondary waves can have much shorter wavelengths than the primary \Revised{waves} and, as a result, they can have much larger damping rates (due to radiative diffusion).  
Systems in which nonlinear interactions are important may therefore dissipate tidal energy much more rapidly than the linear theory estimates. 
Indeed, in the case of solar-type binaries, the linear theory estimates yield dissipation rates that are too small by a factor of $\ga 100$ ($\Q \sim 10^8-10^{10}$; \citealt{Terquem:1998, Goodman:1998, Ogilvie:2007}). 
This may indicate that nonlinear processes are playing an important role in these systems.

For a planet with mass $M_p\ga 3 \Mj (P/\trm{day})^{-0.1}$ orbiting a solar-type star, the primary \Revised{waves reach} such large amplitudes near the stellar center that \Revised{they overturn} the background stratification and \Revised{break} \citep{Barker:2010, Barker:2011b}. 
In this {\it strongly} nonlinear regime, the primary \Revised{waves deposit} nearly all of \Revised{their} energy and angular momentum in a single group travel time through the star. 
The tidal dissipation rate therefore equals the energy flux of the initial, linearly driven primary \Revised{waves.}
The three-dimensional numerical simulations of wave breaking by \citet{Barker:2011b} yield $\Q \simeq 10^5 (P/1\trm{ day})^{2.8}$ for $M_p\ga 3 \Mj$ and a solar-type star. 
This corresponds to an inspiral time of $\approx 1\trm{ Gyr}$ for a $3\Mj$ planet in a 2 day orbit.

For a planet with mass $0.5 \la M_p/\Mj \la 3$, the primary \Revised{waves do not, in general,} break.  
Nonetheless, they are sufficiently nonlinear that \Revised{they excite} many secondary waves near the stellar center. 
In this {\it weakly} nonlinear regime, the primary \Revised{waves only deposit} a fraction of \Revised{their} energy and angular momentum in a single group travel time. 
The value of that fraction, which determines the rate of tidal dissipation, depends on the detailed interaction between the primary \Revised{waves} and the sea of secondary waves. 
The aim of our study is to calculate this interaction (and its saturation) in the weakly nonlinear regime.  
Similar types of analyses have been carried out in the context of the $r$-mode instability in spinning neutron stars \citep{Brink:2005, Bondarescu:2009}.

This paper is structured as follows. 
In \S~\ref{s:formalism} we describe the formalism we use to study the weakly nonlinear tidal interactions and present the equations of motion for our mode decomposition. 
In \S~\ref{s:simulation} we describe how we construct our networks of interacting modes and our method for integrating the coupled equations of motion.  
In \S~\ref{s:mode dynamics} we present a pedagogical discussion of how different mode networks behave. 
The main results of our calculations are presented in \S~\ref{s:orbital decay}, with particular emphasis on the tidal evolution of known exoplanetary systems. 
Finally, in \S~\ref{s:conclusions} we summarize our results and describe some of the limitations of our analysis that can serve as directions for future work.

\section{Formalism}\label{s:formalism}

We are interested in calculating the orbital evolution of hot Jupiters due to tidal dissipation within the host star.  
We assume that the planet's orbit is circular, as is the case for most of the observed hot Jupiters \citep{Udry:2007,Ogilvie:2014}. 
If the system is also sufficiently old so that the planet's rotation is synchronous with the orbit \citep{Storch:2014,Barker:2014}, then there is no tidal dissipation within the planet.

The tide raised by the planet excites a variety of oscillation modes within the star.
Here we limit our analysis to solar-type hosts and focus on the excitation of resonant $g$-modes due to linear and (weakly) nonlinear forces.  
Because the orbital period of a hot Jupiter is much shorter than the rotational period of a solar-type star, the $g$-modes are not strongly modified by Coriolis forces and we therefore neglect the star's rotation. 

\subsection{Equations of Motion}

We calculate the orbital evolution using the formalism developed in WAQB for studying tides in close binary systems in which weakly nonlinear wave interactions are important (see also \citealt{Schenk:2001,VanHoolst:1994}). 
We now briefly summarize the method and refer the reader to  WAQB for a more detailed discussion.  

The equation of motion for the Lagrangian displacement $\gv{\xi}(\gv{r},t)$ of the stellar fluid at position $\gv{r}$ and time $t$ relative to the unperturbed background is
\begin{equation}\label{e:eq_of_motion}
	\rho \ddot{\gv{\xi}} = \gv{f}_1\left[\gv{\xi}\right]+\gv{f}_2\left[\gv{\xi},\gv{\xi}\right]+\rho \gv{a}_{\rm tide},
\end{equation}
\noindent
where $\rho$ is the background density, $\gv{f}_1$ and $\gv{f}_2$ are the linear and leading-order nonlinear restoring forces, 
\begin{equation}
	\gv{a}_{\rm tide}=-\grad U - \left(\gv{\xi}\cdot\grad\right)\grad U
\end{equation}
\noindent
is the tidal acceleration, and $U$ is the tidal potential. We include only the dominant $l=2$ tidal harmonic and since we assume that the orbit is circular, 
\begin{equation}\label{e:tidal_potential}
	U(\gv{r},t)=- \epsilon \omega_0^2 r^2 \sum_{m=-2}^2 W_{2m} Y_{2m}(\theta, \phi)e^{-im\Omega t},
\end{equation}
\noindent
where $\epsilon = (\Mp/M)(R/a)^3$, $\omega_0=(GM/R^3)^{1/2}$ 
is the dynamical frequency of a star with mass $M$ and radius $R$, $\Mp$ is the planet mass, $a$ and $\Omega$ are the orbital semi-major axis and frequency, and $W_{20}=-(\pi/5)^{1/2}$, $W_{2\pm2}=(3\pi/10)^{1/2}$, $W_{2\pm1}=0$.
We solve Equation (\ref{e:eq_of_motion}) using the method of weighted residuals in which we expand the six-dimensional phase space vector as
\begin{equation}\label{e:eig_expansion}
	\begin{bmatrix} \gv{\xi}(\gv{r},t) \\ \partial_t \gv{\xi}(\gv{r},t) \end{bmatrix}
	= 
	\sum_{\alpha} q_\alpha(t) \begin{bmatrix} \gv{\xi}_\alpha(\gv{r}) \\ i\omega_\alpha \gv{\xi}_\alpha(\gv{r}) \end{bmatrix},
\end{equation}
\noindent
where $\alpha$ labels a linear eigenmode with eigenfunction $\gv{\xi}_\alpha$, eigenfrequency $\omega_\alpha$, and amplitude $q_\alpha(t)$. 
The sum over $\alpha$ runs over all mode quantum numbers and frequency signs to allow both a mode and its complex conjugate. We normalize the eigenmodes as
\begin{equation}
	E_0 \equiv \frac{GM^2}{R}=2\omega_\alpha^2\int d^3x \rho \, \gv{\xi}_{\alpha}^\ast \cdot \gv{\xi}_{\alpha},
\end{equation}
\noindent
so that a mode with dimensionless amplitude $|q_\alpha|=1$ has energy $E_0$.  
Plugging Equation (\ref{e:eig_expansion}) into Equation (\ref{e:eq_of_motion}), adding a linear damping term, and using the orthogonality of the eigenmodes leads to a coupled, nonlinear amplitude equation for each mode
\begin{multline}\label{e:amp_eqn}
	\dot{q}_\alpha + (i\omega_\alpha + \gamma_\alpha ) q_\alpha =  \\ 
	i\omega_\alpha \left[U_\alpha(t) + \sum_\beta U_{\alpha\beta}^\ast(t) q_{\beta}^\ast  + \sum_{\beta\gamma} \kappa_{\alpha\beta \gamma}^\ast q_{\beta}^\ast q_\gamma^\ast\right],
\end{multline}
\noindent
where 
\begin{subequations}
	\begin{align}
		U_\alpha(t)                = &-\frac{1}{E_0}\int d^3x \rho \, \gv{\xi}_\alpha^\ast \cdot \grad U,\label{e:Ualpha}\\
		U_{\alpha\beta}(t)         = &-\frac{1}{E_0}\int d^3x \rho \, \gv{\xi}_\alpha \cdot\left(\gv{\xi}_{\beta} \cdot \grad\right)\grad U,\\
		\kappa_{\alpha\beta\gamma} = &\frac{1}{E_0}\int d^3x\, \gv{\xi}_{\alpha}\cdot\gv{f}_2\left[\gv{\xi}_\beta, \gv{\xi}_\gamma\right].
	\end{align}
\end{subequations}
\noindent
The coefficient $\gamma_\alpha$ is the linear damping rate of the mode, $U_\alpha$ and $U_{\alpha \beta}$ represent the linear and nonlinear tidal force, and $\kappa_{\alpha \beta \gamma}$ represents the three-mode coupling.  

\subsection{Expressions for the Coefficients}\label{s:properties of the modes}

We consider the dynamics of high-order, adiabatic $g$-modes within a solar-type main sequence star. 
These modes are restored by buoyancy and propagate between inner and outer turning points determined by the locations at which $\omega_\alpha\simeq N(r)$, where $N$ is the  Brunt-V\"ais\"al\"a buoyancy frequency \citep{Aerts:2010}.  
The inner turning point is very close to the stellar center ($r_{\alpha,\rm inner}/R \simeq 10^{-3} \left(P_{\alpha}/\mathrm{day}\right)^{-1}$) and the outer turning point is near the radiative-convective interface at $\simeq 0.7R$. 
Individual modes are described by the quantum numbers ($l$, $m$, $n$),  where $l$ is the spherical degree,  $m$ is the azimuthal order, and $n$ is the radial order. 
Since the $g$-modes we consider are all very high-order ($n\ga50$), their properties are well approximated by the asymptotic WKB expressions given in WAQB. 
Using a 5 Gyr old solar model from the EZ code \citep{Paxton:2004}, we find 
\begin{subequations}
	\begin{align}
		\omega_\alpha &\simeq 7 \frac{l_\alpha}{n_\alpha}  \omega_0, \label{e:omega_approx} \\
		\gamma_\alpha &\simeq 2\times10^{-11} \Lambda_\alpha^2\left(\frac{\omega_0}{\omega_\alpha}\right)^2 \omega_0, \label{e:gamma_approx}
	\end{align}
\end{subequations}
\noindent
where $\Lambda_\alpha^2 = l_\alpha(l_\alpha+1)$.
The dominant linear damping mechanism \Revised{of the high-order $g$-modes} is radiative diffusion of the temperature fluctuations that accompany the mode density perturbations \citep{Terquem:1998, Goodman:1998}. 
Shorter wavelength modes therefore have larger damping rates. 

By plugging Equation (\ref{e:tidal_potential}) into Equation (\ref{e:Ualpha}), we can express the linear driving coefficient $U_\alpha$ in terms of the dimensionless linear overlap integral
\begin{subequations} \label{e:forcing}
	\begin{align}
		I_\alpha & = \frac{1}{MR^2}\int d^3x \rho\, \gv{\xi}_{\alpha}^\ast \cdot\grad\left(r^2 Y_{2m}\right) \\
 		         & \simeq 2.5\times 10^{-3}\left(\frac{\omega_\alpha}{\omega_0}\right)^{11/6},
	\end{align}
\end{subequations}
\noindent
where the numerical expression assumes $l_\alpha=2, m_\alpha=m$ and is accurate for mode periods $P_\alpha \ga 0.3\trm{ day}$ (Figure 11 of WAQB).
Low-order, $l=2$ $g$-modes have large $I_\alpha$ but $\omega_\alpha \gg \Omega$; they comprise the quasi-static response of the fluid (the equilibrium tide).  
High-order, $l=2$ $g$-modes have small $I_\alpha$ but can nonetheless have large linear amplitudes if $\omega_\alpha \simeq 2\Omega$; they comprise the resonant response of the fluid (the dynamical tide). 

The three-mode coupling coefficient $\kappa_{\alpha \beta \gamma}$ is symmetric under the interchange of mode indices.  
Angular momentum conservation leads to the following angular selection rules for the three modes: (i) $l_\alpha+l_\beta+l_\gamma$ must be even, (ii) $m_\alpha+m_\beta+m_\gamma= 0$, and (iii) the triangle inequality, $\left| l_\alpha-l_\beta\right| \leq l_\gamma \leq l_\alpha + l_\beta$.
We focus on the parametric instability involving three-mode interactions between a high-order ``parent" $g$-mode and a pair of  high-order ``daughter" $g$-modes whose summed frequency nearly equals the parent's frequency.
For such a triplet, the coupling is strongest in the stellar core,  where the Lagrangian displacements of the modes peak. For a solar-type star (Appendix A in WAQB)
\begin{equation}\label{e:kappa_approx}
	\kappa_{\alpha \beta \gamma} \simeq 2\times10^3\, \left(\frac{T}{0.2}\right) \left(\frac{P_\alpha}{1\trm{ day}}\right)^2,
\end{equation}
\noindent
where $P_\alpha$ is the period of the parent mode and $T\approx 0.1 -1$ is an angular integral that depends on each mode's $l$ and $m$.  
The coupling occurs mostly near the parent's inner turning point $r_{\alpha, \rm inner}$ and scales as $P_\alpha^2$ because the parent's displacement there varies as $\xi_\alpha \sim  r_{\alpha, \rm inner}^{-2} \sim P_\alpha^{2}$.

Although the equilibrium tide amplitude is large, its three-mode coupling cancels significantly with nonlinear tidal driving $U_{\alpha \beta}$  (WAQB; see also \citealt{Venumadhav:2014}). 
As a result, for a hot Jupiter system, the nonlinear dynamics are dominated by three-mode coupling to the dynamical tide;  the energy dissipated in the equilibrium tide is small by comparison.
\footnote{\Revised{Turbulent dissipation of the equilibrium tide within the convection zone yields $\Q\sim 10^8-10^9$ \citep{Penev:2011}, which is much larger than the $\Q$ we find due to nonlinear damping of the dynamical tide (\S~\ref{s:orbital decay}).}}  We therefore restrict our mode networks to parent modes that comprise the dynamical tide response of the star (i.e., linearly resonant parents) and ignore the equilibrium tide response and nonlinear tidal driving. 

Finally, we assume that only linearly resonant modes (parents) have non-zero linear tidal forcing $U_\alpha$. 
This is justified because the linear forcing coefficient $U_\alpha$ is much smaller for daughter modes and their driving is far off resonance.  
Its secular effect will therefore be negligible compared to the resonant three-mode interactions.  
Ignoring such forcing allows us to adopt a convenient change of coordinates that significantly speeds up the integration of the amplitude equations (see \S~\ref{s:integration}).  

\subsection{Instantaneous Orbital Decay Time-scale}

The energy of the stellar modes is $E_\ast/E_0 = \sum q_\alpha q_\alpha^\ast + (1/3) \sum k_{\alpha\beta\gamma}(q_\alpha q_\beta q_\gamma + c.c)$. 
The rate of energy loss within a solitary star is therefore
\begin{eqnarray} \label{e:disp}
	\frac{\dot{E}_\ast}{E_0} & = &\sum_\alpha \left( \dot{q}_\alpha q_\alpha^\ast + c.c \right) + \sum_{\alpha\beta\gamma} k_{\alpha\beta\gamma} \left( \dot{q}_\alpha q_\beta q_\gamma + c.c \right) \nonumber\\ 
	             & = & -2\sum_\alpha \gamma_\alpha q_\alpha q_\alpha^\ast + \sum_{\alpha\beta\gamma} k_{\alpha\beta\gamma} \gamma_\alpha \left( q_\alpha q_\beta q_\gamma + c.c \right), \nonumber\\  
	             \dot{E}_\ast & \approx&  -2 \sum_\alpha \gamma_\alpha E_\alpha,
\end{eqnarray}
\noindent
where we substituted the equations of motion (Equation \ref{e:amp_eqn}) for the mode amplitudes and neglected the terms from the three-mode couplings because they are much smaller than $E_\alpha=q_\alpha q_\alpha^\ast E_0$.
This is the rate at which energy is dissipated within the star by radiative diffusion.

The dissipation of tidally excited stellar modes removes energy from the orbit, and the orbit therefore decays.
We assume that the only dissipation in the system is due to the linear damping of waves excited within the star. 
Although the rotational energy of a synchronized planet increases as the orbit decays, this change is small compared to the corresponding change in orbital energy. 
Similarly, the energy in the excited stellar modes themselves may change with orbital period, but this also is a small effect \Revised{(see Appendix \ref{s:time averaging})}.

Because $|\dot{E}_\ast/E_{\rm orb}| \ll \Omega$, where $E_{\rm orb}=-GM\Mp/2a$ is the orbital energy, we model the back-reaction on the orbit as a steady decrease in $E_{\rm orb}$ of quasi-Keplerian circular orbits. 
The timescale of the instantaneous, orbital energy decay is then given by
\begin{equation}\label{e:instant tau}
	\T = \frac{E_{\rm orb}}{\dot{E}_\ast},
\end{equation}
\noindent
at each \Porb, and we can compute a corresponding time-averaged decay time-scale
\begin{equation}\label{e:tau}
	\aveT = \frac{a}{\left|\dot{a}\right|}=\frac{E_{\rm orb}}{\langle \dot{E}_\ast \rangle},
\end{equation}
\noindent
where $\langle \dot{E_\ast} \rangle$ is the time-averaged energy dissipation rate with the average spanning several resonance peaks (if  $\aveT \propto a^n$ then the ``inspiral time" into the star will be $\aveT /n$). 
We describe our method of time-averaging in Appendix \ref{s:time averaging}.
Using the language of linear tidal theory, \aveT\ is often parameterized in terms of the star's tidal quality factor (\citealt{Goldreich:1966}; see also \citealt{Jackson:2008})
\begin{equation}
        \Q = 7.5\times10^{6} \left(\frac{\aveT}{\mathrm{Gyr}}\right) \left(\frac{\Mp}{\Mj}\right) 
                                                \left(\frac{\Porb}{\mathrm{day}}\right)^{-13/3}  \label{e:Q}
\end{equation}
\noindent
where the expression assumes a circular orbit about a solar-type star and ignores dissipation within the planet. 
Although \Q\ is often taken to be a constant and fundamental property of the body, in general it  depends on the companion mass, orbital frequency, and the tidal harmonics $(l, m)$.  

\section{Building and integrating the mode networks}\label{s:simulation}

In the absence of nonlinear three-mode interactions, the energy of a linearly driven parent mode $\alpha$ is (Equation 29 in WAQB)
\begin{equation}\label{e:Elin}
	\frac{E_{\rm lin}}{E_0} = \left| q_{\alpha, \rm lin} \right|^2 = \frac{ \omega_\alpha^2 U_\alpha^2 } {\Delta_\alpha^2 + \gamma_\alpha^2},
\end{equation}
\noindent
where $\Delta_\alpha = \omega_\alpha - m_\alpha \Omega$ is the linear detuning.  
The parent is unstable to nonlinear three-mode interactions if there exists a pair of daughter modes $\beta$ and $\gamma$ such that  $E_{ \rm lin} \ga E_{\rm thr}$, where the threshold energy is (see Appendix \ref{s:3mode equilib derivation})
\begin{equation}\label{e:Ethr}
	\frac{E_{\rm thr}}{E_0} = \frac{1}{4\kappa_{\alpha\beta\gamma}^2}\left(\frac{\gamma_b \gamma_\gamma}{\omega_\beta \omega_\gamma}\right)\left[1+\left(\frac{\Delta_{\beta \gamma}}{\gamma_\beta+\gamma_\gamma}\right)^2\right].
\end{equation}
\noindent
Here $\Delta_{\beta \gamma}=\omega_\beta+\omega_\gamma+m_\alpha\Omega$ is the nonlinear detuning of the daughter pair.  
Daughter pairs with smaller $|\Delta_{\beta \gamma}|$ and larger $|\kappa_{\alpha\beta\gamma}|$ (i.e., stronger nonlinear coupling) yield smaller $E_{\rm thr}$ and are more readily unstable. 
In a three-mode system, unstable daughters with small initial amplitude undergo a phase of exponential growth at a rate
\begin{equation}\label{e:s3}
	\Gamma_{\rm 3md} \approx 2 \Omega \left|\kappa_{\alpha\beta\gamma}\right| \sqrt{E_{ \rm lin}/E_0}.
\end{equation}
\noindent
Eventually, the daughters reach an energy comparable to or greater than the parent's and the system reaches a nonlinear equilibrium (see \S~\ref{s:parents and daughters} and Appendix \ref{s:3mode equilib derivation}).  

For the tide raised by even a $0.1\, \Mj$ companion in a 3 day orbit, there are $\sim 10^3$ daughter pairs for which $E_{\rm thr}<E_{ \rm lin}$ (see \S~\ref{s:selecting three-mode triples}). 
In \S~\ref{s:mode dynamics}, we systematically explore the dynamics of large multi-mode, multi-generation systems.  
In brief, we find that the parent drives many of the unstable daughters to large amplitudes and these daughters, in turn, drive granddaughters to large amplitudes, and so on.  
The total number of potentially unstable modes and the number of couplings is larger than the number we can integrate on a computer in a reasonable time ($\sim 10^4$ and $\sim 10^5$, respectively). 
The issue then is whether we can reliably calculate the total tidal dissipation rate with a mode network that contains only a subset of the potentially unstable modes.  
We will present evidence in \S~\ref{s:mode dynamics} that this is possible but we must build our networks carefully and systematically.  

In \S\S~\ref{s:selecting three-mode triples} and \ref{s:selecting collective sets}, we describe how we build networks consisting of sets of three-mode couplings (i.e., sets of triplets) and collective  couplings, respectively.  And in \S~\ref{s:integration}, we describe our method for integrating the coupled mode amplitude equation (Equation \ref{e:amp_eqn}).

\subsection{Building Three-mode Networks}\label{s:selecting three-mode triples}

Although there are many daughter pairs with $E_{\rm thr}<E_{\rm lin}$, we show in \S~\ref{s:mode dynamics} that pairs with low $E_{\rm thr}$ dominate the dynamics of large multi-mode systems.  
We find that if we gradually increase the size of our networks by adding pairs with progressively higher $E_{\rm thr}$, the system converges to a dissipation rate $\dot{E_\ast}$ that does not change significantly as we add even more modes.  
We must also include a sufficient number of generations (at least parents, daughters, and granddaughters) in order to obtain convergent results.  
Therefore, to build our mode networks, we comprehensively search the mode parameter space and construct, for each generation, a complete list of pairs ranked by $E_{\rm thr}$.

In order to carry out our search, we use the expressions for $\omega$, $\gamma$, and $\kappa$ (Equations \ref{e:omega_approx}, \ref{e:gamma_approx}, and \ref{e:kappa_approx}) to solve for $E_{\rm thr}$.  
For a given parent mode $\alpha$, we first find the local minima of $E_{\rm thr}$ in the daughter parameter space $\{(n_{\beta}$, $l_{\beta}$, $m_{\beta}$), ($n_{\gamma}$, $l_{\gamma}$, $m_{\gamma})\}$. 
In general, $E_{\rm thr}$ is minimized approximately where the sum in quadrature of $\Delta_{\beta \gamma}$ and $\gamma_\beta + \gamma_\gamma$ is minimized (modulo the angular selection rules and a relatively weak dependence on the angular integral $T$). 
Daughters with higher $l$ have smaller $\Delta_{\beta \gamma}$ (because they are more densely spaced in frequency) but larger $\gamma$ (because $\gamma\sim l^2$); the regions of small $E_{\rm thr}$ therefore occur where these two countering effects are balanced.  
After finding the local minima, we expand our search around those minima and find pairs with progressively higher $E_{\rm thr}$. 
Because $\gamma\sim l^2$, at high enough $l$ the damping dominates detuning, and $E_{\rm thr}$ increases with increasing $l$.  
We truncate our search upon reaching an $l_{\rm max}$ such that \Revised{$E_{\rm thr}>E_{\rm lin}$} (i.e., a stable triplet). 
In practice, we find that for parent-daughter coupling, the dissipation $\dot{E}_\ast$ is dominated by the 10--100 lowest $E_{\rm thr}$ triplets.  

\begin{figure}
	\begin{center}
		\includegraphics[width=1.0\linewidth]{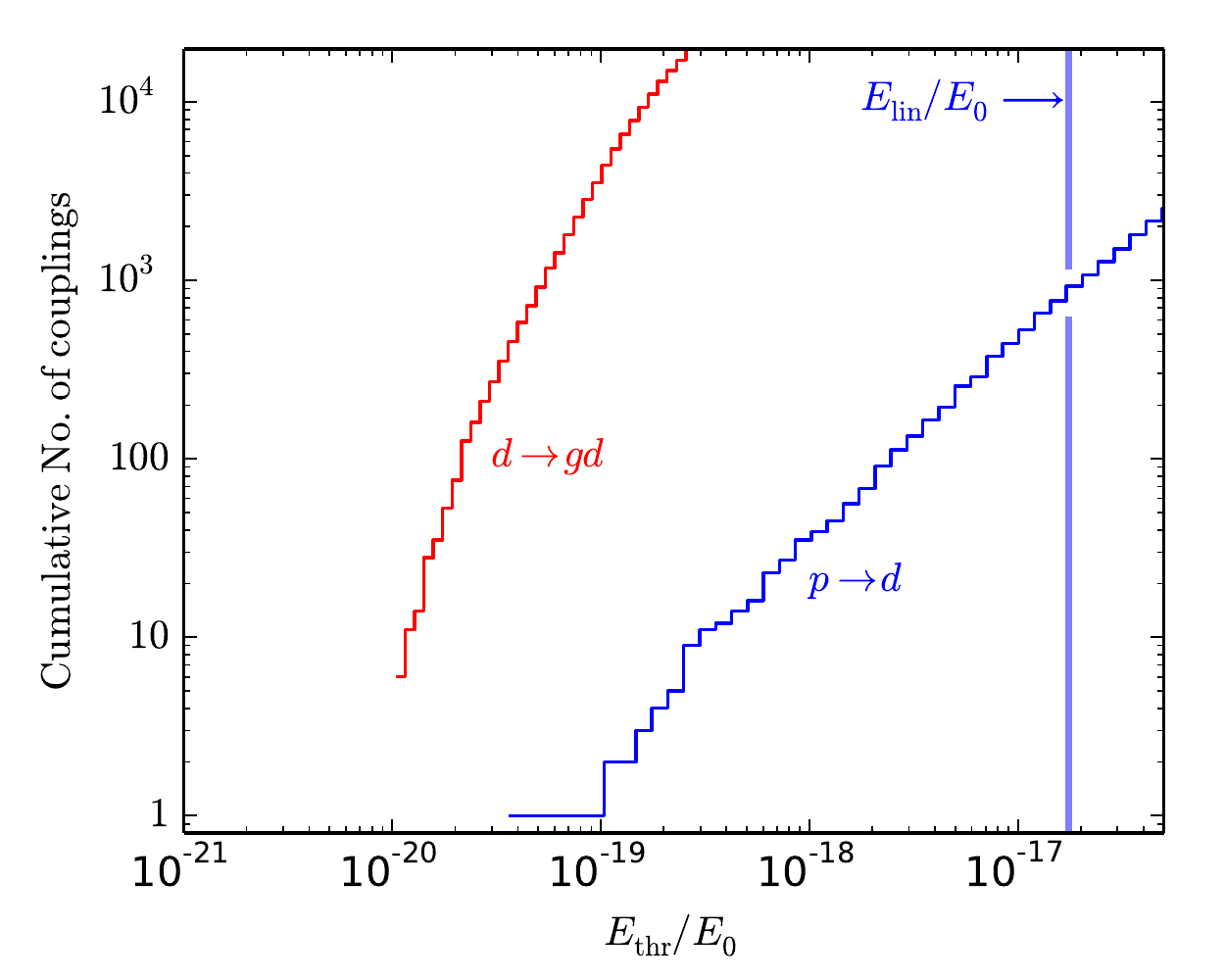}
	\end{center}
	\caption{
		Cumulative distribution of $E_\mathrm{thr}$ for $M_p=0.1\Mj$ and $\Porb\simeq3$ days (257928 sec). 
        (blue) Couplings from parents to daughters.
        The vertical blue line corresponds to the parent's linear energy $E_{\rm lin}$.
        (red) Couplings from daughters to granddaughters. 
	}
	\label{f:Athr distribution}
\end{figure}

Figure \ref{f:Athr distribution} shows the distribution of $E_\mathrm{thr}$ for parent-daughter coupling assuming a 0.1 \Mj\ companion in an orbit near three days.  
There are a few pairs with very low $E_\mathrm{thr}$ because they just happen to have particularly small $\Delta_{\beta \gamma}$ despite having $l\la 3$.
There is a much larger sample of $\sim 10^3$ pairs that have larger $\Delta_{\beta\gamma}$ and/or $\gamma$ which still yield $E_{\rm thr} \ll E_{ \rm lin}$. 

We carry out a similar search when we consider the coupling of daughters to granddaughters.   We show the distribution of $E_\mathrm{thr}$ for daughter-granddaughter coupling in Figure \ref{f:Athr distribution}. 
The $E_{\rm thr}$ of the most unstable daughter-granddaughter triplets is much smaller than the $E_{\rm thr}$ of the most unstable parent-daughter triplets (i.e., the red curve in Figure \ref{f:Athr distribution} is far to the left of the blue curve).
Because $\kappa_{\alpha\beta\gamma}\propto \omega_{\alpha}^{-2}$ and $\Delta_{\beta\gamma}/\omega_\alpha \propto \omega_{\alpha}$, for low $E_{\rm thr}$ pairs we find $E_{\rm thr}\propto (\Delta_{\beta\gamma}/\kappa_{\alpha\beta\gamma}\omega_\alpha)^2\propto \omega_\alpha^6$. 
The factor of two decrease in frequency with each generation therefore means that $E_{\rm thr}$ decreases by $\sim 2^6$ (for a full discussion, see Appendix \ref{s:breaking thr}). 
Physically, $E_{\rm thr}$ decreases because lower frequency modes (i) penetrate deeper into the core where Lagrangian displacements are larger, and (ii) are more densely spaced in frequency and therefore can have smaller detunings.   
As a result, each generation is evermore susceptible than its predecessor to three-mode instabilities. 
This has important implications for the dynamics of large multi-mode, multi-generation systems, as we describe in \S~\ref{s:mode dynamics}. 

\subsection{Building Collective Networks}\label{s:selecting collective sets}

For high-order $g$-modes, the frequency spacing between neighboring modes is $|\Delta \omega| \sim \omega/n \ll \omega$.  
Therefore, if a pair of daughters is resonantly excited by a parent, there is a good chance that neighboring modes will also be resonantly excited by that same parent.  The dynamics of such a system can be very different from  that of a simple three-mode system; in particular, the daughters can grow as a single, collective unit with growth rates that are much higher than the three-mode case (see WAQB and Appendix \ref{s:collective stability}).  

To appreciate why collective sets can grow so quickly, consider a simplified system in which a single parent mode $\alpha$ is coupled to $N$ daughter modes that are closely-spaced neighbors in $(l,n)$ space.  
A study of the dynamics of such a system reveals that the modes all oscillate nearly in phase with each other. 
The equations of motion for each of the $N$ daughter modes can thus be approximated as
\begin{eqnarray}
	\dot{q}_\beta + \left(i\omega_\beta +\gamma_\beta\right) q_\beta & = & i\omega_\beta \sum_\gamma \kappa_{\alpha\beta\gamma}^\ast q_\alpha^\ast q_\gamma^\ast \nonumber \\ 
		                                                         & \simeq & i\omega_\beta N \kappa_{\alpha\beta\gamma}^\ast  q_\alpha^\ast q_\gamma^\ast.
\end{eqnarray}
\noindent
The dynamics look like the three-mode case, but with an effective coupling coefficient that is $N$ times larger. 
In particular, the instability growth rates (threshold amplitudes) are approximately $N$ times larger (smaller) than the three-mode case.

When building our mode networks, we use separate algorithms to search for collective sets and three-mode sets. 
A simple but incomplete way to build collective sets is to first find a daughter with a frequency nearly equal to half that of the parents and then progressively add neighbors with $\Delta n = \pm1, \pm2, \pm3,\ldots$.
At first, $E_{\rm thr}$ will decrease as more modes are added and $N$ increases. 
However, for large enough $\Delta n$, the detuning of the outer most modes becomes so large that adding more modes does not decrease $E_{\rm thr}$ any further.\footnote{In addition,  the magnitude of $\kappa_{\alpha\beta\gamma}$ becomes small for $\Delta n \ga n_{\alpha}$ because the coupled daughters are no longer spatially resonant with the parent (see Figure 12 in WAQB).}
More detail is provided in Appendix \ref{s:decoupling of large detuning}.
Although this method naturally picks out collective sets (and is similar to the approach described in WAQB), it potentially misses many collectively unstable modes.  
For example, there can be distinct groups of modes that are not nearby neighbors and yet together form a collective set. 
For this reason, we use a more sophisticated method when building collective networks.  We describe this method in Appendix \ref{s:collective}.  

In \S~\ref{s:mode dynamics}, we show that collective sets are excited and initially grow much more rapidly than three-mode sets. 
However, when the entire network ultimately reaches its nonlinear equilibrium and saturates, we find that the collective sets do not \Revised{alter $\dot{E}_\ast$ significantly}.  
We therefore find that we can accurately calculate $\dot{E}_\ast$ with networks that include only three-mode sets.

\subsection{Integration Method}\label{s:integration}

We integrate the amplitude equation (Equation \ref{e:amp_eqn}) for each mode of a network using an adaptive step-size \Revised{4$^{th}$-5$^{th}$ order Runge-Kutta integrator}. 
Our integrations take advantage of a convenient change of coordinates, also described in \cite{Brink:2005}.
The integration step size is limited by the fastest frequency in the equations. 
Because the linear and nonlinear forcings all involve resonant interactions\footnote{For reasons described in \S~\ref{s:properties of the modes}, we assume that only the linearly resonant parent modes have a non-zero linear tidal forcing $U_\alpha$.}, the linear and nonlinear detunings ($\Delta_\alpha$ and $\Delta_{\beta\gamma}$) are all small ($\ll \Omega$). In fact, the fastest time scale in these equations is typically the natural frequency $\omega_\alpha$ of each mode. 
By changing coordinates to $x_\alpha = q_\alpha e^{i \omega_\alpha t}$, we can remove these frequencies from the equations of motion at the cost of adding a slowly varying time-dependent term to each three-mode coupling.
This increases the typical integration step size by approximately the ratio of $\omega_\alpha$ to the detuning ($\approx 10^{2} - 10^{4}$).

In order to further speed-up the integrations,  we parallelize across multiple CPUs.
We achieve this using standard parallelization techniques, with care taken to equally distribute the amount of work across each CPU.
For example, the computation of $\dot{x}_\alpha$ scales with the number of couplings included for that mode.
Therefore, when we parallelize the computation of $\dot{x}_\alpha$ by splitting modes among processes, we attempt to divide modes into sets with equal numbers of couplings, rather than equal numbers of modes.
We test several different parallelization methods, including an implementation using Python's subprocess module, Python's multiprocessing module, and a Python wrapper for OpenMPI.
All our implementations scale better than $N_\mathrm{CPU}^{-0.8}$, although which implementation is fastest depends on specifics of the hardware.
The Python multiprocessing implementation generally performed best, and parallelized across 15 2.7 GHz Quad-Core AMD Opteron Processors, it takes 
$\simeq 40$ seconds to integrate one of our largest networks ($\simeq 2.4\times10^4$ modes with $\simeq 3\times10^5$ couplings)
through 10 orbital periods.
\begin{figure*}
	\begin{minipage}{0.5\linewidth}
        \includegraphics[width=1.0\columnwidth]{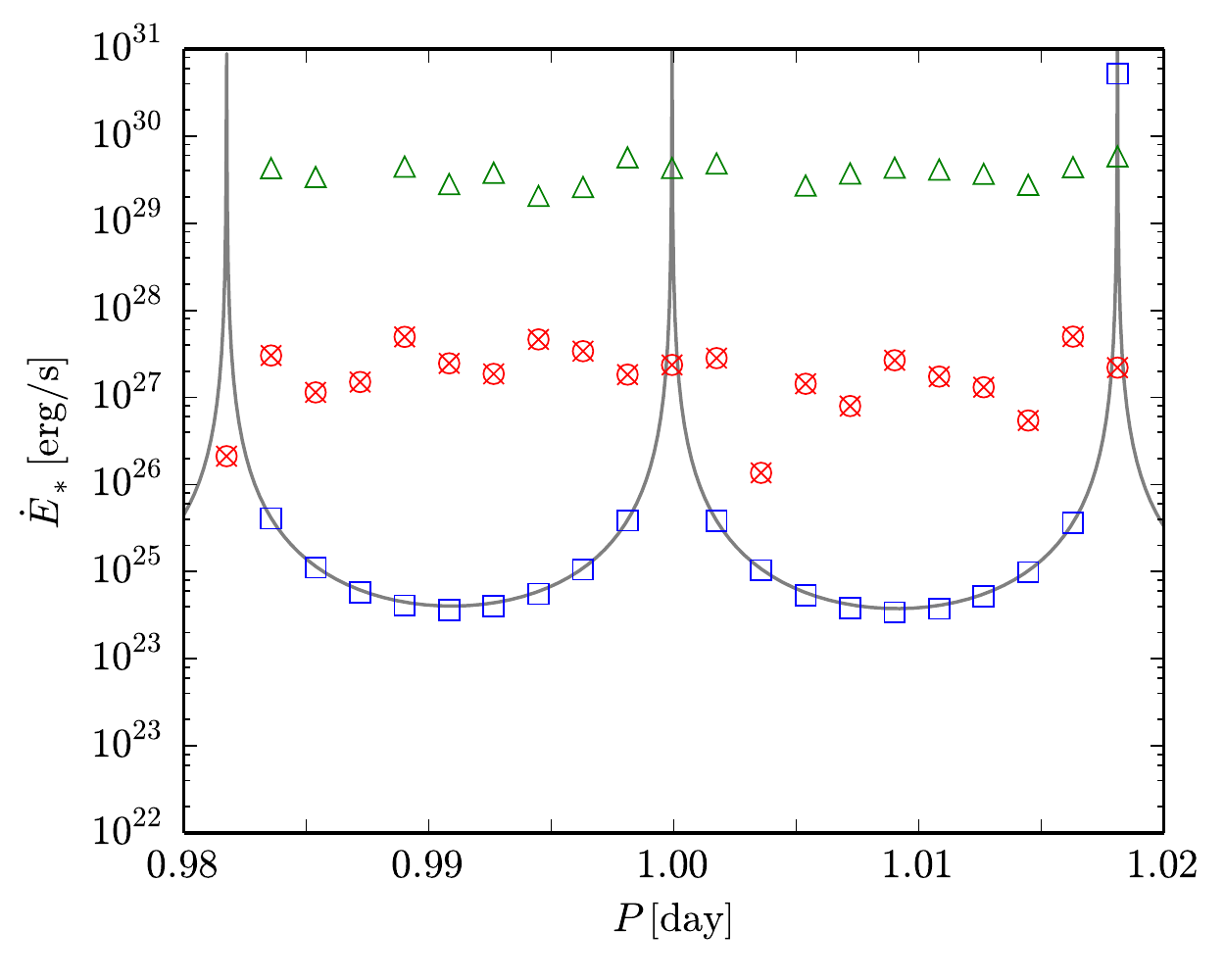}
	\end{minipage}
	\begin{minipage}{0.5\linewidth}
		\includegraphics[width=1.0\columnwidth]{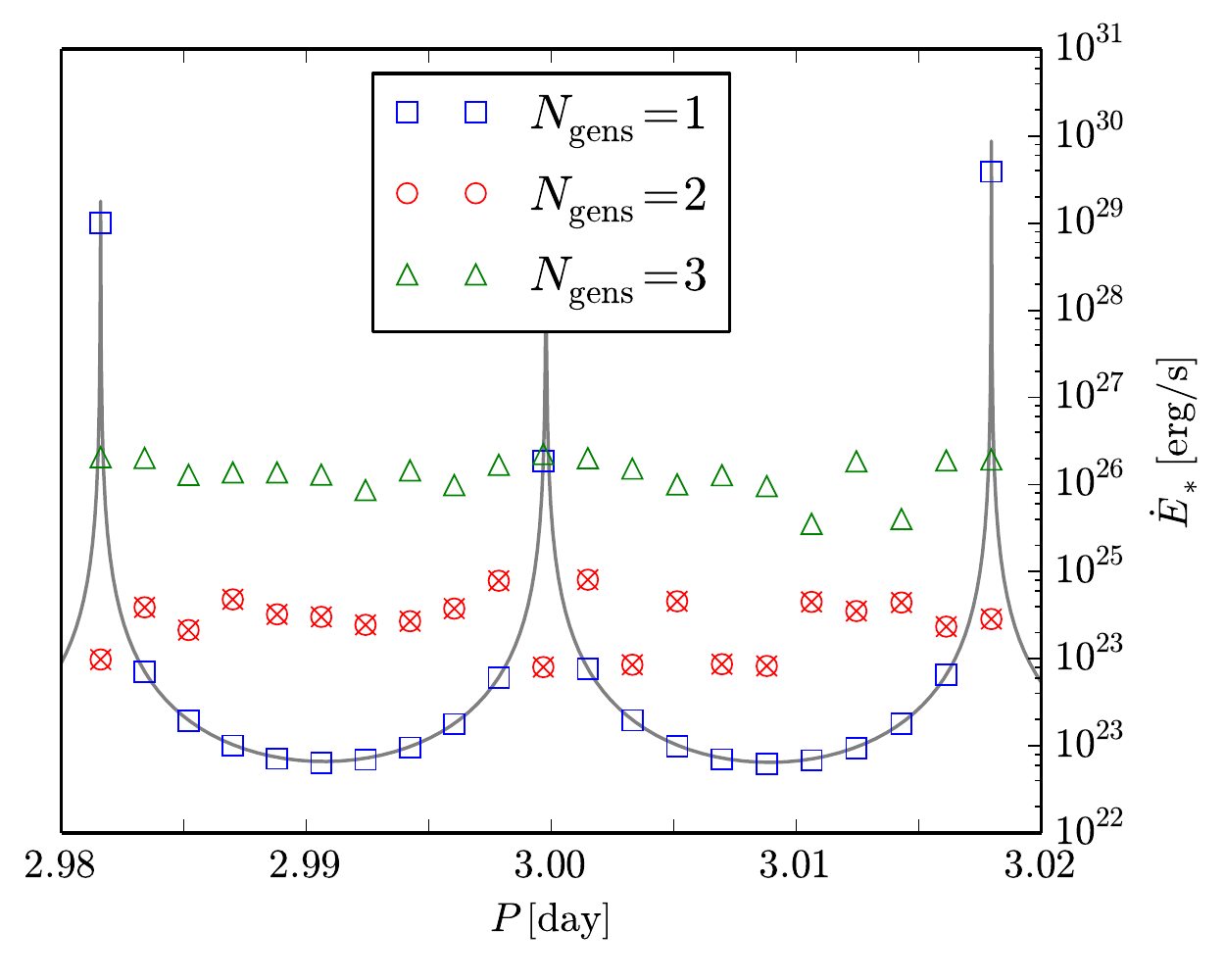}
	\end{minipage}
        \caption{
		Energy dissipation rate $\dot{E}_\ast$ as a function of orbital period $P$ for networks that include different numbers of mode generations $N_{\rm gens}$.  We show results for $M_p=\Mj$ and orbits near $P= 1\trm{ day}$ (left panel) and $P= 3\trm{ days}$ (right panel). The sharp linear resonance peaks occur when the tidal driving frequency is resonant with an $l=2$ $g$-mode of the star.
        (blue squares) Networks with only the ten most resonant parents.
        (red circles) Networks with parents and daughters.
        (green triangles)  Networks with parents, daughters and granddaughters (shown here are the results of our ``reference network"; see \S~\ref{s:saturation summation}).
        (gray lines) The analytic estimate of the steady state dissipation for parent-only networks that contain the 100 most resonant parents.
        (red crosses) The analytic estimate of the steady state dissipation for parent-daughter networks (see Equation \ref{e:aveT_pd}). 
        We do not have an analytic estimate for parent-daughter-granddaughter networks. 
	}
        \label{f:detailed sweep}
\end{figure*}
\section{mode dynamics}\label{s:mode dynamics}

In order to build up intuition for the results from large multi-mode, multi-generation networks, we describe the mode dynamics of increasingly complicated networks. We begin in \S~\ref{s:parents} with a network that consists only of linearly resonant parents (i.e., we ignore all nonlinear couplings) and show that our simulations recover the dissipation rates of standard linear theory.  
In \S~\ref{s:parents and daughters} we couple linearly resonant parents to unstable daughter modes but do not allow the daughters to couple to granddaughters. 
We find that including even just this first generation of nonlinear couplings enhances the dissipation rate by a factor of $\approx 100$ ($\approx 10$) relative to the linear result for $\Porb=1\trm{ day}$ (3 day) and $\Mp = \Mj$. 
In \S~\ref{s:parents, daughters and granddaughters} we allow the daughters to couple to granddaughters and find that this further enhances the dissipation, yielding a rate that is $\approx 10^5$ ($\approx 10^3$) times larger than the linear result for $\Porb=1\trm{ day}$ (3 day) and $\Mp = \Mj$.
We find in \S\S~\ref{s:great granddaughters} and \ref{s:collective networks} that the dissipation rates do not change significantly when we include even more generations (great granddaughters and beyond) and collective sets, respectively,  suggesting that the system has reached a convergent, saturated state. 
In \S~\ref{s:saturation summation} we explore the minimum network size needed to attain such a convergent state.

\subsection{Linear Parents Only}\label{s:parents}

\begin{figure*}
	\begin{center}
	         \includegraphics[width=0.95\linewidth, clip=True, trim=1.1cm 0.0cm 1.1cm 0.0cm]{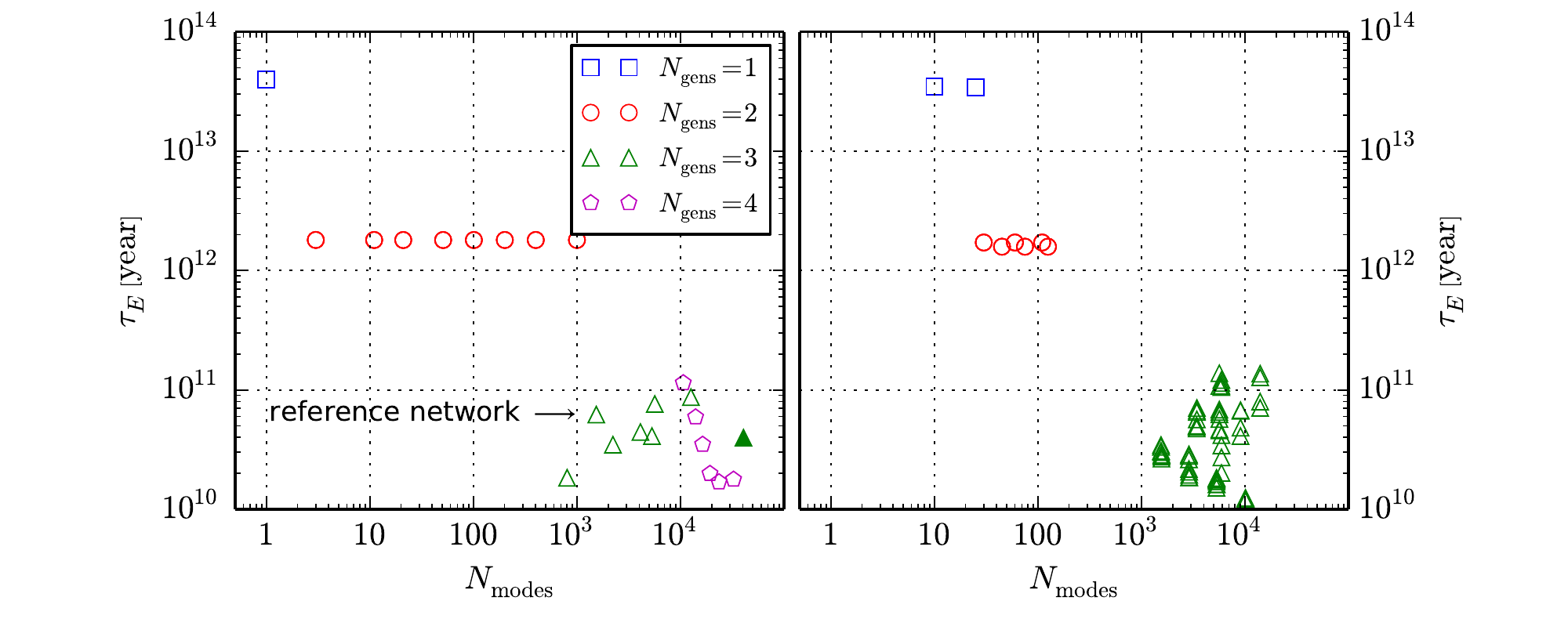}
	\end{center}
        \caption{ 
        Orbital decay timescale \T\ as a function of the number of modes $N_{\rm modes}$ for networks that include different numbers of mode generations $N_{\rm gens}$. 
        We show results for a Jupiter-mass companion orbiting at a period near three days (257928 sec, chosen to be approximately half way between a resonance trough and resonance peak for our stellar model). 
        The left panel networks have a single parent mode and the right panel networks have 10 or 25 parent modes. 
        (blue squares) Networks with only parents.
        (red circles) Networks with parents and daughters.
        (green triangles) Networks with parents, daughters, and granddaughters. \Revised{The filled triangle corresponds to a reference network with collective granddaughter modes added.}
		(purple pentagons) Networks with parents, daughter, granddaughters, and great-granddaughters.
		The structure of the networks with only a single parent (left panel) are as follows:
        	$N_{\rm gens}=2$ networks range from one daughter pair up to 500 daughter pairs. 
            $N_{\rm gens}=3$ networks mostly correspond to 10 daughter pairs and either 25, 50 (our reference network; \S~\ref{s:saturation summation}), 75, 150, or 200 granddaughter pairs per daughter, although there are networks with 50 and 200 daughter pairs, each with 50 granddaughter pairs per daughter. 
            $N_{\rm gens}=4$ networks are all extensions of our reference network, adding 10, 25, 50, \Revised{100, 200, or 500} great-granddaughter pairs per granddaughter.
         The structure of the multi-parent networks (right panel) are as follows: either 10 or 25  parent modes; 0, 10, 25, or 50 daughter pairs per parent; and either 0, 50, 100, or 200 granddaughter pairs per daughter.
	}
        \label{f:multiparent}
\end{figure*}

If we include only linearly driven parents in the network, then $|\dot{E}_\ast| = 2\sum \gamma_\alpha \left(E_\alpha\right)_{ \rm lin}$, where the parent linear energy $\left(E_\alpha\right)_{ \rm lin}$ is given by Equation (\ref{e:Elin}).  
The dissipation is typically dominated by the most linearly resonant parent, although other modes can contribute if no single mode is particularly resonant.
In Figure \ref{f:detailed sweep}, we show $\dot{E}_\ast$ due to the ten most resonant parents over a small range in orbital period.  
In the absence of nonlinear interactions, the orbit evolves rapidly through the sharp resonance peaks where $\dot{E}_\ast$ is large.  
As we describe in Appendix \ref{s:time averaging}, the time average dissipation rate is the sum of the instantaneous $\dot{E}_\ast$ weighted by the amount of time spent at that period (see also \citealt{Goodman:1998}).  
In the left panel of Figure \ref{f:multiparent}, we show \T\ (Equation \ref{e:instant tau}) due to the single most resonant parent assuming $\Porb\simeq3\trm{ day}$, $\Mp=\Mj$.
An analytic calculation using Equations (\ref{e:gamma_approx}) and (\ref{e:Elin}), and assuming $\Delta_\alpha \sim \omega_\alpha / 2 n_\alpha \gg \gamma_\alpha$, yields (see Appendix \ref{s:Tlin})
\begin{equation}
	\aveT_{\rm lin} \simeq 1.4\times10^{12} \left(\frac{\Mp}{\Mj}\right)^{-1} \left(\frac{\Porb}{\mathrm{day}}\right)^{3}\trm{ yr},
\end{equation}
\noindent
which translates to
\begin{equation}
    Q^\prime_{\ast,\rm lin} \simeq 1.1 \times 10^{10} \left(\frac{\Porb}{\mathrm{day}}\right)^{-4/3}.
\end{equation}
\noindent
This is in good agreement with our numerical integrations.
The $\Mp$ dependence of $\aveT_{\rm lin}$ is due to the linear forcing coefficient $U_\alpha$ and the dependence on $\Porb$ is due to a combination of $\gamma_\alpha$, $\Delta_\alpha$, and $U_\alpha$.  
We will show that when we include nonlinear interactions, the instantaneous decay time has a dramatically different magnitude and scaling with $\Mp$ and $\Porb$. 

The right panel of Figure \ref{f:multiparent} shows the results for the same parameters as the left panel, but for networks with either the 10 or 25 most resonant parents. 
Despite including these additional modes, the instantaneous decay time is nearly identical in both cases.  
More generally, we find that including multiple parents has very little effect on the total dissipation (even for networks with nonlinear interactions) as long as $\Porb\la 4 \trm{ days}$.  
This is because, for $\Porb\la 4 \trm{ days}$, the parent mode spacing is sufficiently sparse that the most resonant parent typically has a much larger $E_{\rm lin}$ than the neighboring parents, and it therefore dominates the dynamics (Appendix \ref{s:2daughter Nparent}).

\subsection{Parents and Daughters}\label{s:parents and daughters}

\begin{figure}
	\begin{center}
	        \includegraphics[width=1.0\columnwidth, clip=True, trim=0.2cm 0.10cm 0.00cm 0.75cm]{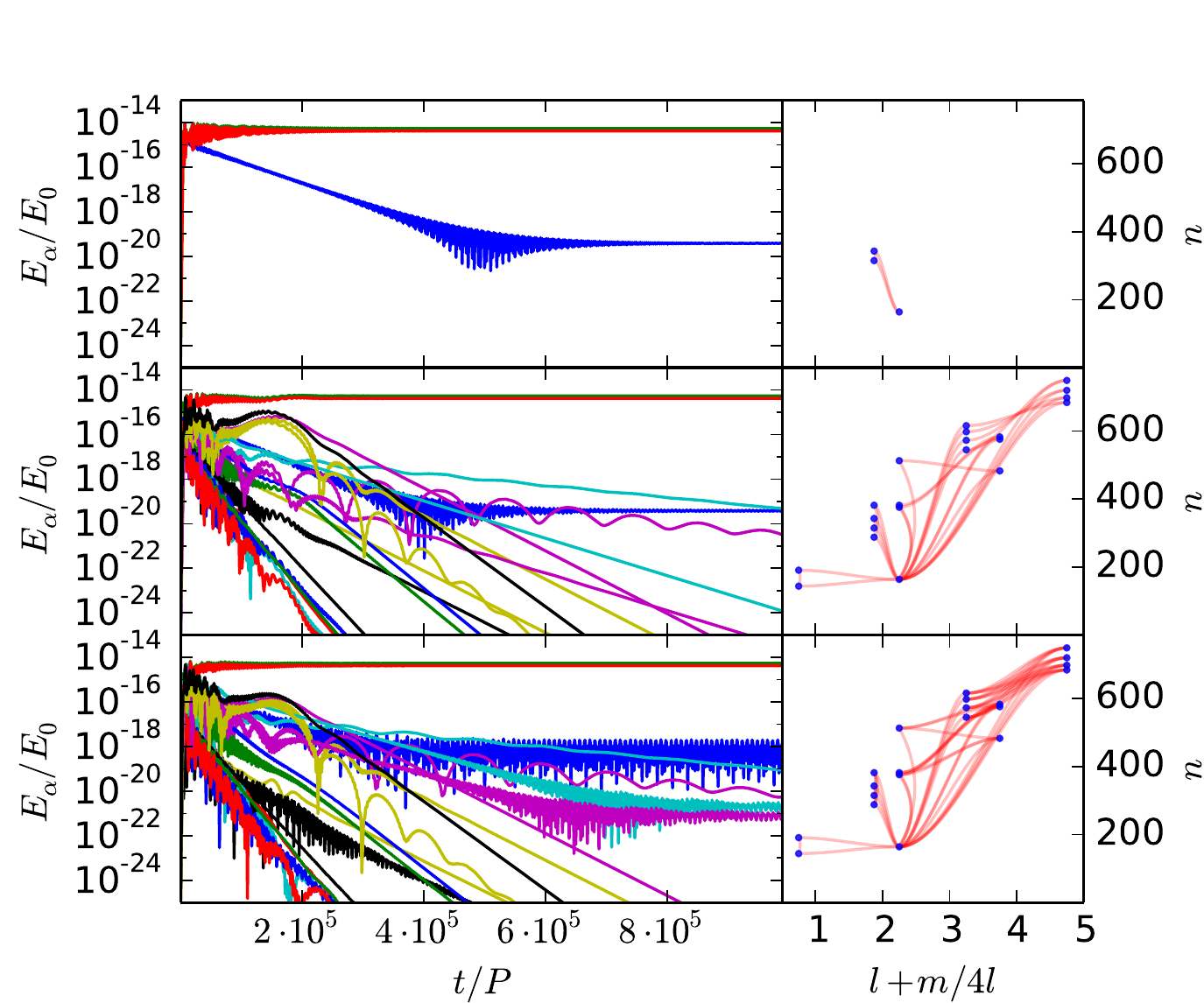}
	\end{center}
        \caption{
		Parent-daughter networks ($N_{\rm gens}=2$) with different structures. 
		The left panels show the  energy $E_\alpha$ of each mode as a function of time.  In each of these panels, the parent is the blue line and is initially at $E_{\rm lin}$ but ultimately settles into a nonlinear equilibrium at an energy $E\ll E_{\rm lin}$.
        The right panels show the coupling diagrams in the $n$-$l$ plane, with circles representing the included modes and line connections indicating the coupling structure.
		(top) A parent and the lowest $E_\mathrm{thr}$ daughter pair. 
		(middle) A parent and the ten lowest $E_\mathrm{thr}$ daughter pairs with each daughter mode couple to only one other daughter mode. 
		(bottom) A parent and the ten lowest $E_\mathrm{thr}$ pairs with all allowed couplings between daughters.
	}
        \label{f:g1 examples}
\end{figure}

In Figure \ref{f:g1 examples} we show the mode dynamics of networks that include daughters (but not granddaughters) coupled to a linearly resonant parent.  
The top panel shows a simple three-mode system involving a parent coupled to only its lowest $E_{\rm thr}$ daughter pair.  
Initially, the daughters are at small energy and the parent is at its linear energy $E_{\rm lin}$.  
Because $E_{\rm lin}> E_{\rm thr}$, the system is unstable and the daughters undergo a rapid initial growth at the rate given by Equation (\ref{e:s3}). 
Eventually the system reaches a nonlinear equilibrium in which the parent has energy $E_{\alpha} = E_{\rm thr}$ and the daughters have energy $E_{\beta,\gamma}\simeq |U_\alpha/2\kappa_{\alpha\beta\gamma}|E_0$  (see Appendix \ref{s:3mode equilib derivation} and WAQB). 

The middle panel of Figure \ref{f:g1 examples} shows the same parent now coupled to the ten lowest $E_{\rm thr}$ daughter pairs.  
Because $E_{\rm lin} > E_{\rm thr}$ for all ten pairs, initially all the daughters grow.  
However, eventually the parent energy drops to the minimum $E_{\rm thr}$ and only the lowest threshold daughter pair remains excited; the other daughters decay due to linear damping.  
The nonlinear equilibrium of this system is therefore equivalent to the three-mode network shown in the top panel.

The network shown in the middle panel of Figure \ref{f:g1 examples}  assumes that each of the ten triplets only share a parent.  
If the triplets also share daughters (e.g., daughter $a$ couples to daughter $b$ and daughter $c$), then the dynamics can be more complicated.  
Such a network is shown in the bottom panel of Figure \ref{f:g1 examples} and illustrates how such additional couplings parasitically excite other daughters.\footnote{This is a form of nonlinear inhomogeneous driving and is described in WAQB.} 
\citealt{OLeary:2014} consider a similar mechanism in order to explain the odd resonances observed in the KOI-54 light curve.  
Despite these additionally excited modes, the lowest $E_{\rm thr}$ pair still dominates the dissipation.

In Figure \ref{f:multiparent}, we show \T\ for networks that include only parent-daughter couplings ($N_{\rm gen}=2$) assuming $\Porb\simeq 3\trm{ day}, \Mp=\Mj$.   
We find that at this period parent-daughter coupling decreases \T\ by a factor of $\sim 10$ relative to the linear result.  
Numerically, both \T\ and \aveT\ are nearly independent of the number of daughter modes in the parent-daughter networks because the daughter pair with the lowest $E_{\rm thr}$ dominates the dynamics and dissipation. 
The other daughters, while excited, do not reach significant amplitudes and therefore have little effect.  
Even for large numbers of modes, parent-daughter systems behave much like those shown in the middle and bottom panels of Figure \ref{f:g1 examples}.  
We therefore find that \aveT\ is well-approximated by the analytic calculation that assumes only one parent and its single lowest $E_{\rm thr}$ daughter pair (see Appendix \ref{s:Tlin} and Figure \ref{f:frequency sweeps})
\begin{equation}\label{e:aveT_pd}
        \aveT_{\textrm{p-d}} \simeq 2.0  \times10^{11} \left(\frac{\Porb}{\mathrm{day}}\right)^{19/6}\trm{ yr},
\end{equation}
\noindent
assuming $l=1$ daughters and
\begin{equation}
        Q^\prime_{\ast,\textrm{p-d}} \simeq 1.5 \times 10^{9} \left(\frac{\Mp}{\Mj}\right) \left(\frac{\Porb}{\mathrm{day}}\right)^{-7/6}
\end{equation}
\noindent
The agreement between the numerical result for parent-daughter networks containing multiple daughters and the above three-mode estimate is further illustrated in Figure \ref{f:detailed sweep}.  
The open circles show the numerically computed $\dot{E}_{\ast}$ of a parent-daughter network consisting of the $\simeq 20$ lowest $E_{\rm thr}$ daughter modes.  
Each of these open circles is covered by an ``x", which represent the analytically  computed $\dot{E}_{\ast}$ assuming only the minimum $E_{\rm thr}$ daughter pair.

We will now see, however, that the dynamics are much more complicated when granddaughters are included, with many more modes excited to significant amplitudes.

\subsection{Parents, Daughters, and Granddaughters}\label{s:parents, daughters and granddaughters}

\begin{figure}
	\begin{center}
		\includegraphics[width=1.0\columnwidth, clip=True, trim=0.75cm 0.50cm 1.50cm 0.30cm]{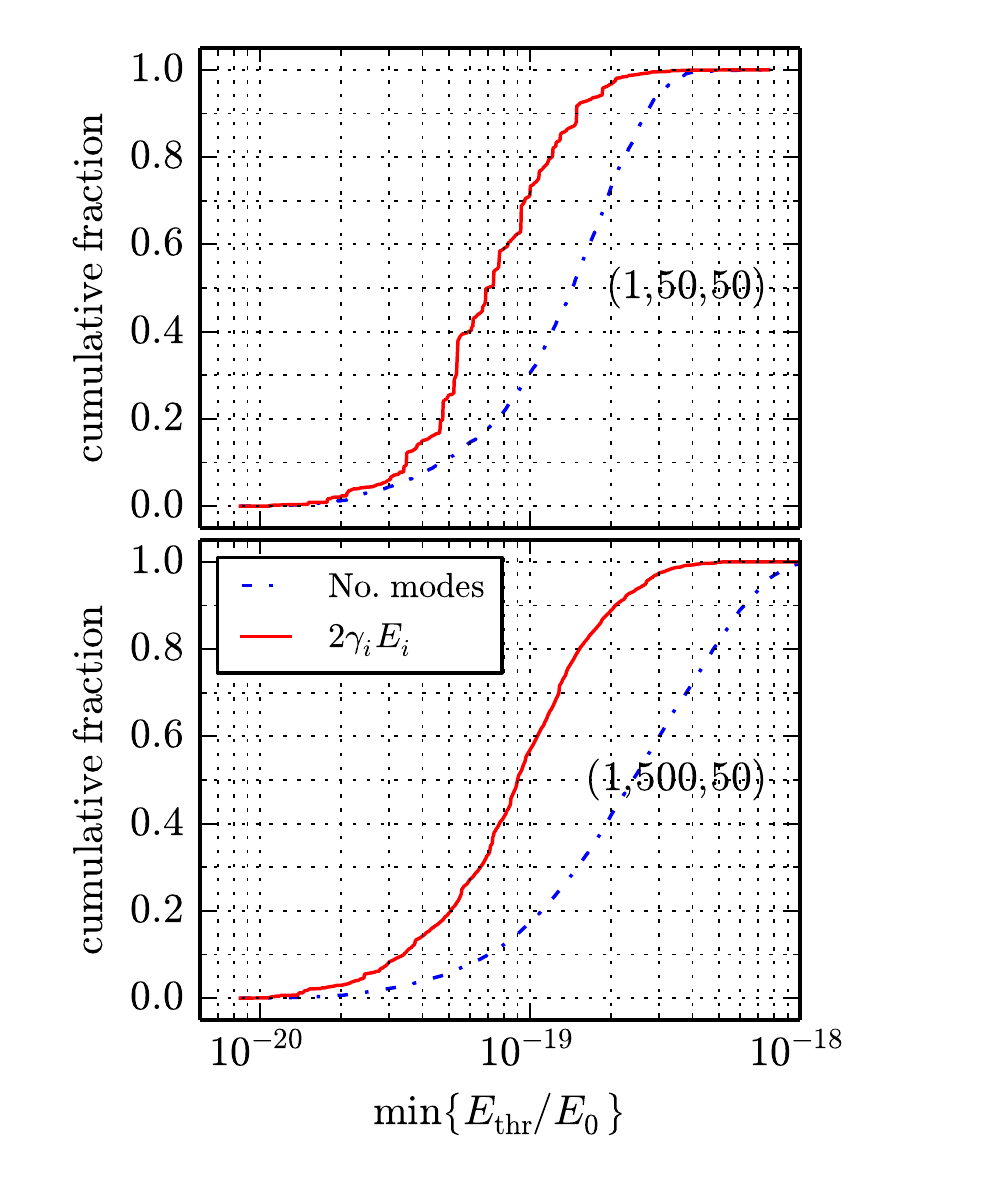}
	\end{center}
	\caption{
		\Revised{
			Cumulative distributions of the number of modes included in the network (blue) and the energy dissipation rate of the modes (red) as a function of each mode’s $E_{\rm thr}$. 
			For granddaughter modes coupled to more than one daughter, we take their minimum $E_{\rm thr}$.  
			(top) a network consisting of one parent, 50 daughter pairs, and 50 granddaughter pairs per daughter. 
			(bottom) a network consisting of one parent, 500 daughter pairs, and 50 granddaughter pairs per daughter.  
			Approximately 90\% of the energy dissipation is due to the first 50\%  (40\%) of modes ordered by $E_{\rm thr}$ for the smaller (larger) network.
		}
	}
	\label{f:stacked hist}
\end{figure}

In the absence of granddaughter couplings, the lowest $E_{\rm thr}$ daughter pair settles into a nonlinear equilibrium at an energy $E_{\beta,\gamma}\simeq|U_\alpha /2\kappa_{\alpha \beta\gamma}|E_0$.  
However, as discussed in  \S~\ref{s:selecting three-mode triples}, there are many granddaughter pairs that are unstable to such high energy daughters (see Figure \ref{f:Athr distribution}).  
The parent-daughter solutions of the previous section are therefore unstable and never realized.   

For very small networks that include granddaughters we sometimes observe periodic limit cycles.   However, for even slightly larger networks with more complicated coupling topologies, the limit cycles begin to take on a more chaotic appearance.   And for the very large networks that we find yield convergent dissipation results ($\ga 10^3$ modes), the dynamics cease to display any clear limit cycle behavior over long time scales and instead show persistent large amplitude fluctuations involving many excited modes (Figure \ref{f:reference examples}).  

We can roughly understand the behavior of these networks using intuition from simple three-mode systems. 
Initially, an unstable, linearly driven parent excites daughters to large energy.  
The daughters drain energy from the parent and the parent's energy drops.  
However, the daughters then excite granddaughters and the daughters' energy drops.  
The daughters no longer drain enough energy from the parent and the parent begins to recover due to linear driving. 
The rising parent excites the daughters again and the cycle restarts.

Unstable granddaughters have lower frequencies and, in general, higher $l$ than the parents and daughters.  
They therefore often have much smaller radial wavelengths (i.e., much larger $n$) and thus much larger linear damping rates ($\gamma \propto n^2$).  
This means that granddaughters can dissipate energy more rapidly than daughters even if they are at a lower amplitude.  

In Figure \ref{f:multiparent} we show \T\ for large networks that include parents, daughters, and granddaughters ($N_{\rm gen}=3$) assuming $\Porb\simeq3\trm{ day}$ and $\Mp=\Mj$.  
We see that networks with granddaughters are more dissipative than parent-daughter only networks and yield $\T \la 10^{-3} \tau_{E,\rm{lin}}$ at $P\simeq 3\trm{ day}$.  
The figure also shows \T\ as a function of the number of modes $N_{\rm modes}$ in the network.  
We find a systematic uncertainty in \T\ associated with the structure of the network.
However, this uncertainty is small compared to the increase in dissipation associated with the inclusion of granddaughter modes.
In this sense, we find that \T\ is not particularly sensitive to the number of granddaughters nor to the details of the network structure as long as the number of granddaughter modes is sufficiently large ($\ga 10^3$). We illustrate this point further when we discuss our reference networks in \S~\ref{s:saturation summation}.

\Revised{
When we build networks with larger $N_{\rm modes}$, we do so by adding modes of increasingly larger $E_{\rm thr}$ (see \S \ref{s:selecting three-mode triples}).  
The fact that \T\ does not change as we increase $N_{\rm modes}$  above $\sim 10^3$ suggests that the lowest $E_{\rm thr}$ modes dominate the energy dissipation and, therefore, that our method for building networks reliably captures the bulk of the dissipation.  
We illustrate this more explicitly in Figure \ref{f:stacked hist}, which shows that the overwhelming majority of the energy is dissipated by the modes with the lowest $E_{\rm thr}$ and that modes with ever larger $E_{\rm thr}$ contribute less and less to the total dissipation.  
This suggests that selecting modes based on their $E_{\rm thr}$ identifies the dynamically relevant couplings and that including enough modes in this way yields convergent results.
}


\subsection{Great Granddaughters and Beyond}\label{s:great granddaughters}

If daughters and granddaughters are excited, what about great granddaughters ($N_{\rm gen}=4$) and so on?  
This seems particularly likely given that $E_{\rm thr}\propto \omega^{-6}$ (see \S~\ref{s:selecting three-mode triples}). 
Indeed, based on our experiments with networks that include up to five generations, we observe that the cascade continues into many generations.  
However, as long as we include enough modes, we find that the total dissipation rate plateaus once we include granddaughters.  
In effect, we do not need to resolve the innermost scales of the energy cascade in order to obtain an accurate estimate of $\dot{E}_\ast$. 

We illustrate this in the left panel of Figure \ref{f:multiparent}, which shows \T\ for networks that go up to $N_{\rm gen}=4$.  
There are dramatic decreases in \T\ when going from just parents ($N_{\rm gen}=1$) to parents and daughters ($N_{\rm gen}=2$), and again when adding granddaughters ($N_{\rm gen}=3$).  However, we see only a slight  decrease in \T\ when we add great-granddaughters ($N_{\rm gen}=4$).  
\Revised{In particular, for sufficiently large $N_{\rm gen}=4$ networks ($N_{\rm modes}\ga 2\times10^4$), we find that \T\ plateaus at a value that is only $\simeq 3$ times smaller than that of our $N_{\rm gen}=3$ reference network at this \Porb.  This suggests that truncating at $N_{\rm gen}=3$ yields reasonably accurate estimates of \T.}


\begin{figure}
        \begin{center}
                \includegraphics[width=1.0\columnwidth, clip=True, trim=0.25cm 0.00cm 1.95cm 0.35cm]{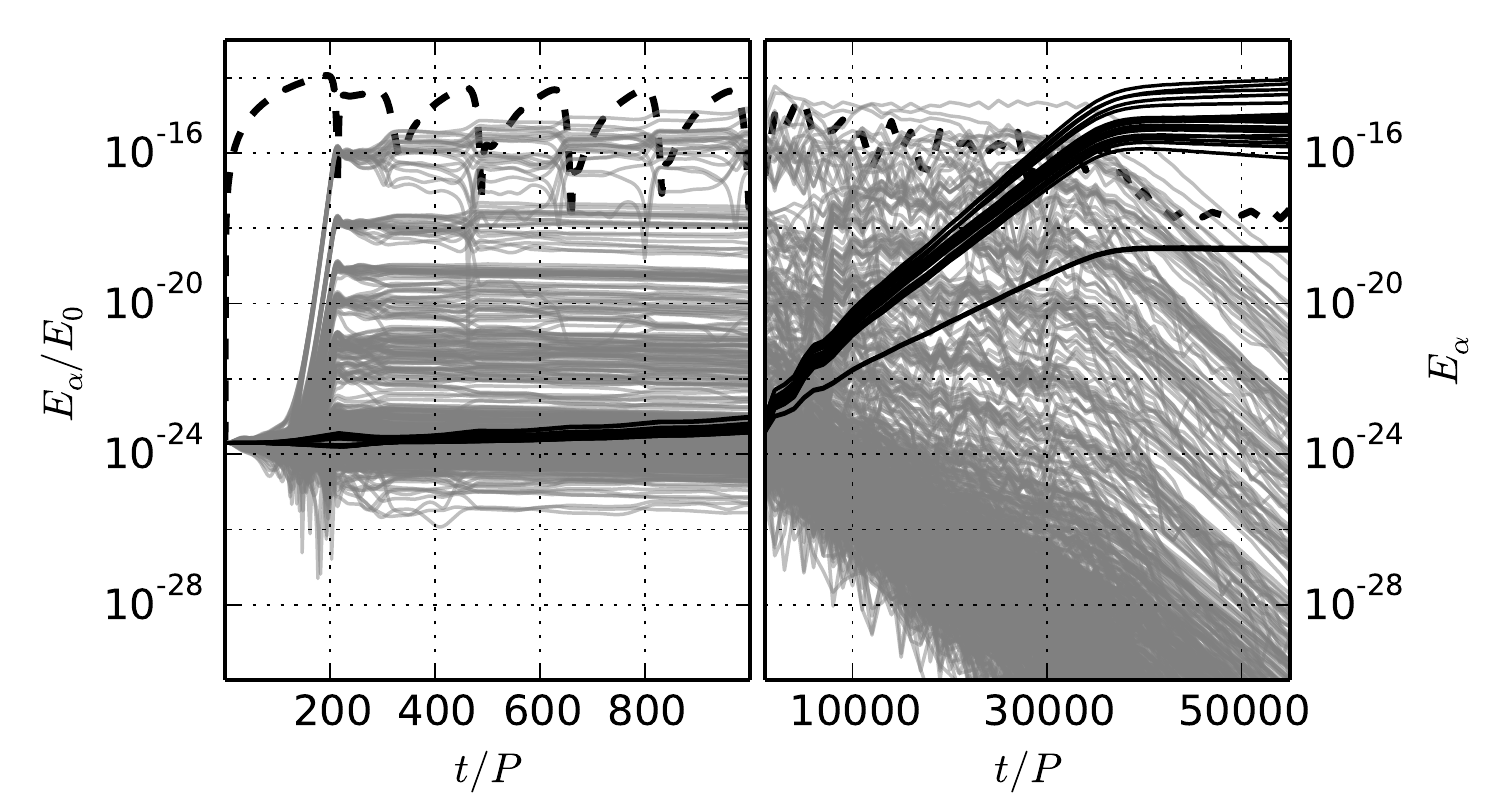}
        \end{center}
        \caption{
              Parent-daughter networks ($N_{\rm gens}=2$) that include collective sets in addition to three-mode sets. \Revised{(black dashed line) the parent mode. (black solid line) daughters selected with three-mode algorithm. (grey solid lines) daughters selected with collective algorithms.} In the left panel we show the dynamics of the system at early times. Several separate sets of collective modes are excited, grow rapidly, and reach significant energies by $t\approx 200 P$. In the right panel we show the dynamics out to much longer times. The slowly growing set of unstable three-mode daughters eventually reach large amplitudes at $t\approx 3\times10^4P$. This drives the parent's amplitude down and the collective modes decay away. We only plot one out of every ten collective modes from this simulation.
        }
        \label{f:collective examples}
\end{figure}

\begin{figure*}[!]
        \begin{center}
                \includegraphics[width=1.0\textwidth, clip=True, trim=1.90cm 0.50cm 1.50cm 0.95cm]{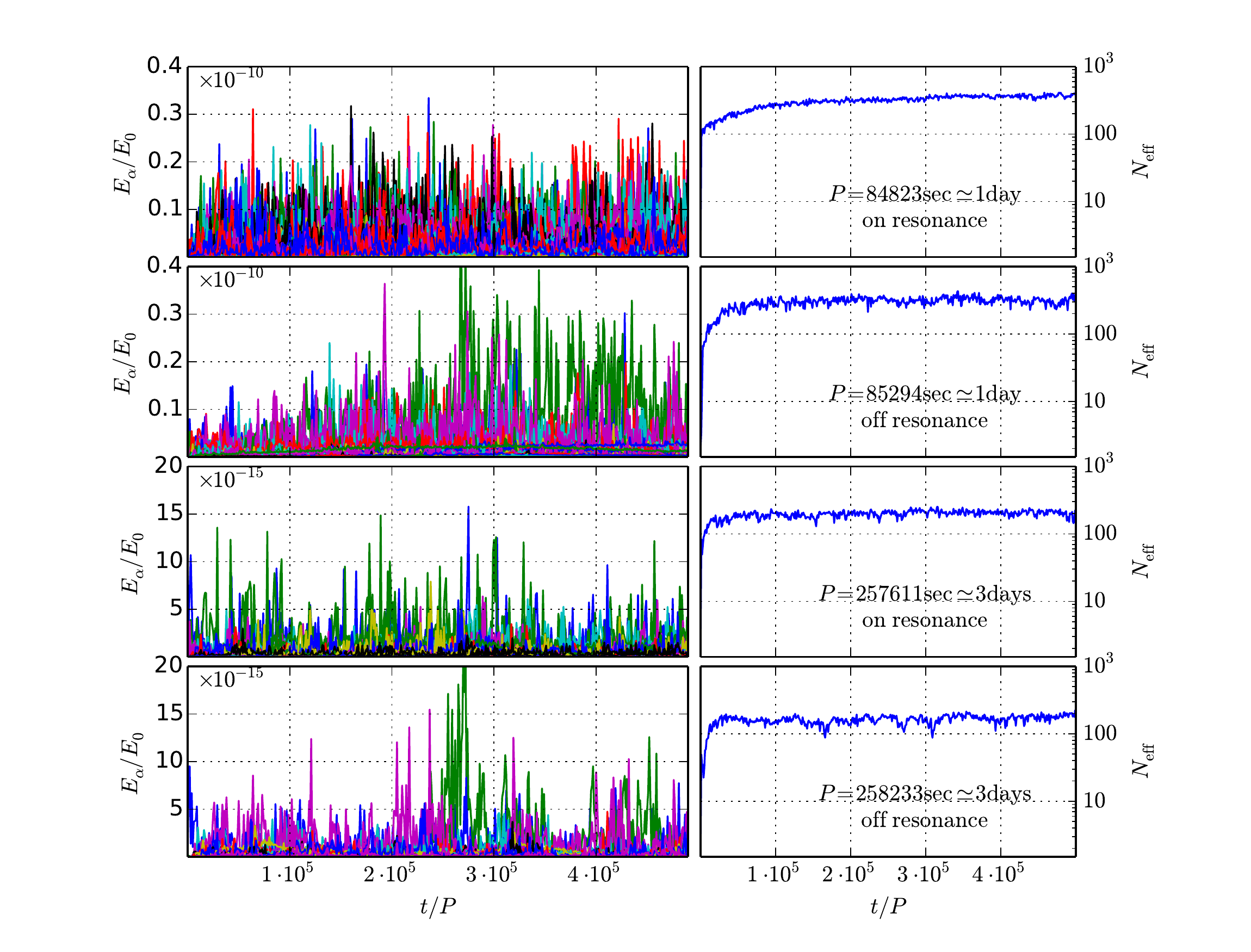}
        \end{center}
        \caption{
                Mode energy $E_\alpha$ (left panels) and effective number of modes $N_{\rm eff}$ (right panels) as a function of time for reference networks with $M_p=\Mj$.  
                (top to bottom) $\Porb\simeq 1$ day on resonance, off resonance, $\Porb\simeq 3$ day on resonance, off resonance.
                For clarity, on the left we plot only one out of every 25 granddaughter modes included in the simulation.
        }
        \label{f:reference examples}
\end{figure*}

\subsection{Dynamics of Collective Networks}\label{s:collective networks}

WAQB showed that sets of modes can be collectively unstable even if each pair within the set is stable by itself. 
We describe algorithmic approaches to identify and select such sets of modes in \S~\ref{s:selecting collective sets} and Appendix \ref{s:collective stability}.
These collective sets can have growth rates that are hundreds of times faster than the separate three-mode growth rates (Equation \ref{e:s3}).  
We illustrate this in the left panel of Figure \ref{f:collective examples}, which shows several collectively unstable sets of modes being rapidly driven to large amplitudes after only a few hundred orbital periods.
However, for $\Porb\la10$ days (see Figure 7 of WAQB), the lowest $E_{\rm thr}$ for individual three-mode triples is lower than the collective stability threshold. 
This means that after the collective modes grow rapidly, the parent's amplitude is still large enough to drive three-mode triples.
The right panel of Figure \ref{f:collective examples} shows that the slowly growing three-mode triples eventually reach large amplitudes and drive the parent below the collective instability threshold.
At that point, all the collective modes ``turn off'' and decay, leaving the steady state predicted by simple three-mode systems.

The network in Figure \ref{f:collective examples} only includes parents and daughters. \Revised{However, because these daughters are unstable to granddaughter interactions (\S~\ref{s:parents, daughters and granddaughters}), the collective modes may not decay forever but instead may saturate at non-trivial amplitudes. In principal, because collective modes can have significantly larger $l$ than the minimum $E_{\rm thr}$ pair and thus larger damping coefficients, they may dissipate energy more rapidly.
Nonetheless, numerical experiments reveal that sufficiently large networks constructed out of only three-mode pairs yield nearly the same $\dot{E}_\ast$  as networks that also include collective excitations. This can be seen in Figure \ref{f:multiparent}; the filled triangle corresponds to the reference network (\S~ \ref{s:saturation summation}) with the addition of collective granddaughters.}
 Because collective networks are expensive to simulate and do not change the calculated $\dot{E}_\ast$, from here on we do not include them in our calculations.
\begin{figure*}
        \begin{minipage}{0.5\linewidth}
                \begin{center}
                        \includegraphics[width=1.0\columnwidth, clip=True, trim=0.30cm 0.05cm 0.18cm 0.35cm]{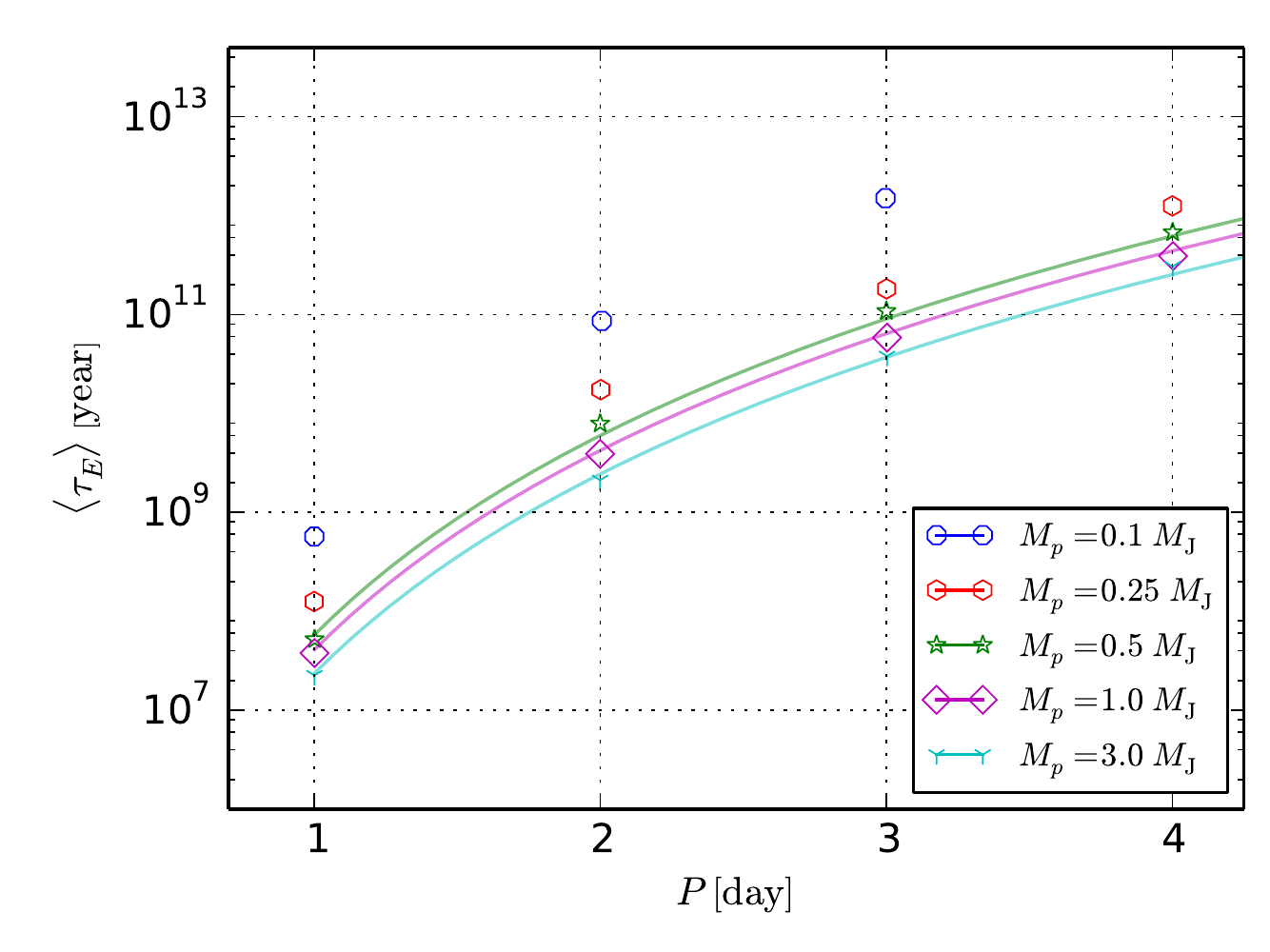}
                \end{center}
        \end{minipage}
        \begin{minipage}{0.5\linewidth}
                \begin{center}
                        \includegraphics[width=1.0\columnwidth, clip=True, trim=0.40cm 0.05cm 0.13cm 0.35cm]{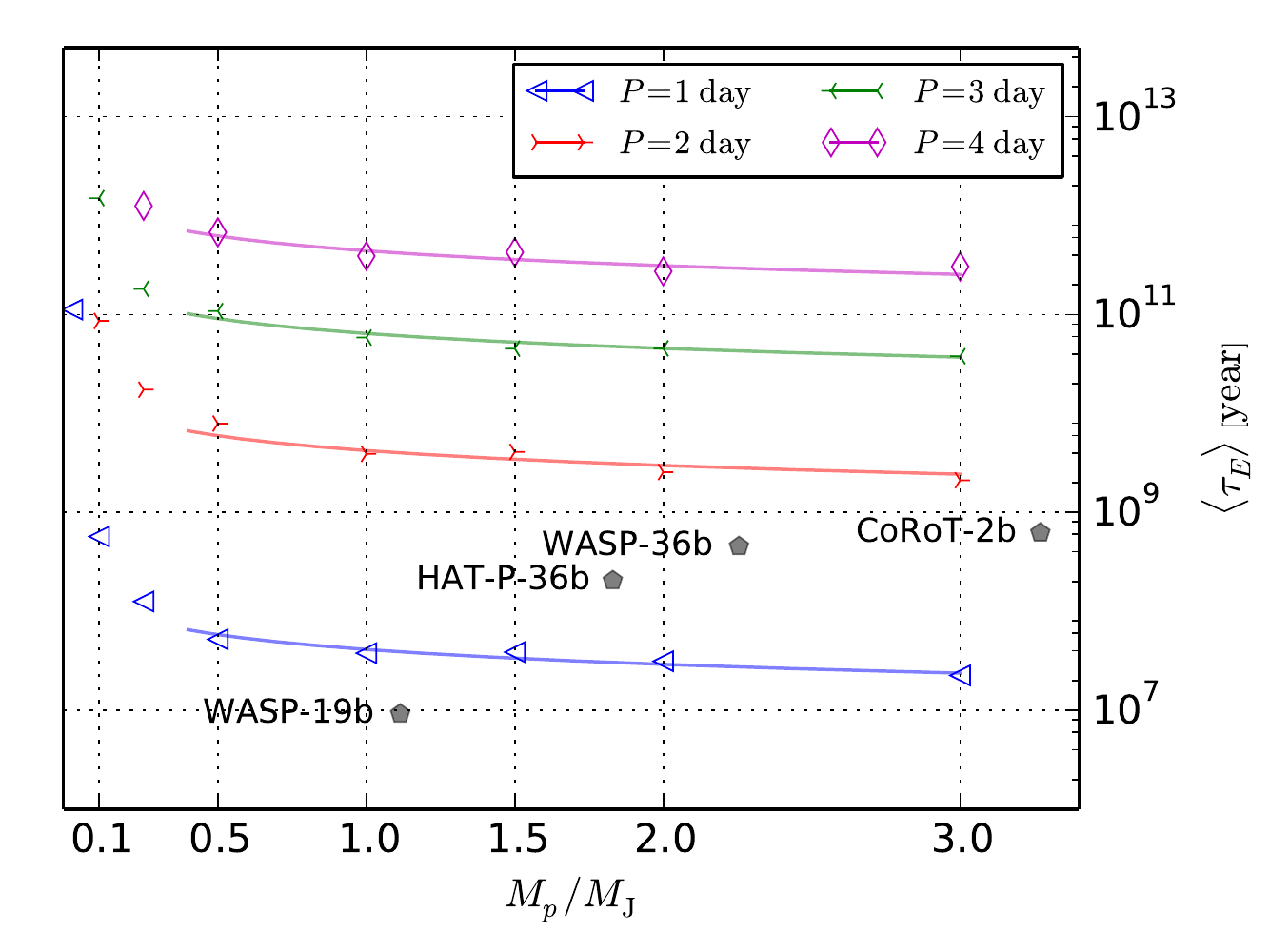}
                \end{center}
        \end{minipage}
        \caption{
                Dependence of \aveT\ on $\Porb$ for fixed values of $\Mp$ (left panel) and the dependence of \aveT\ on $\Mp$ for fixed values of $\Porb$ (right panel).  The solid lines show our analytic fitting formula, which begins to break down for $\Mp \la 0.5 \Mj$.          
}
        \label{f:mass-period decomposition}
\end{figure*}
\subsection{Reference Network Integrations}\label{s:saturation summation}

To summarize, we find that networks with parents, daughters, and granddaughters yield convergent dissipation results as long as they include a large enough number of low $E_{\rm thr}$ daughter and granddaughter modes.  
Moreover, it is not necessary to include collective sets of daughters and granddaughters; although they can modify the dynamics somewhat, the low $E_{\rm thr}$ three-mode sets ultimately \Revised{model the total dissipation well.}

Further numerical experiments reveal that a network consisting of one parent,  its $\simeq 20$ lowest $E_{\rm thr}$ daughters, and $\simeq$1500 low $E_{\rm thr}$ granddaughters is sufficiently large that it yields convergent results while still allowing us to efficiently explore a range of $\Mp$ and $\Porb$ (see caption of Figure \ref{f:multiparent}).  
We use this as our ``reference network" when computing \aveT\ as a function of $\Mp$ and $\Porb$ in \S~\ref{s:orbital decay}.  
That such a network is sufficiently large can be gleaned from the left panel of Figure \ref{f:multiparent}, which shows that the reference network  \T\ is very similar to $N_{\rm gen}=3$ or 4 networks with $N_{\rm modes}\ga 10^4$. 

We demonstrate this further in Figure \ref{f:reference examples}, which shows the mode energy and the effective number of modes participating in the dissipation $N_{\rm eff}$ as a function of time for four different orbital periods.  
We estimate $N_{\rm eff}$ by computing 
\begin{equation}
	N_\mathrm{eff} = \exp\left( - \sum_\alpha p_\alpha \ln p_\alpha \right),
\end{equation}
\noindent
where
\begin{equation}
 p_\alpha = \frac{ \gamma_\alpha q_\alpha q_\alpha^\ast }{\sum_\alpha \gamma_\alpha q_\alpha q_\alpha^\ast }.
\end{equation}
\noindent
This statistic is related to the Shannon entropy and is similar to one used in \citet{Brink:2005}.  
If all modes contribute equally to the dissipation, then $p_\alpha=1/N_{\rm modes}$ for each mode and $N_{\rm eff}=N_{\rm modes}$.  
We see in Figure \ref{f:reference examples} that the dynamics are complicated, with many excited modes.  
We also see that the mode energy and $N_\mathrm{eff}$ increase at shorter $P$ and near linear resonances.  
However, $N_{\rm eff} \ll N_{\rm modes}$ while both the peak energy of the modes and $N_\mathrm{eff}$ remain nearly constant after $\sim$100,000 orbits, indicating that our reference networks are sufficiently large.
\Revised{The behavior of the networks as a whole is what is important here, rather than the dynamics of any individual mode.
Individual modes fluctuate significantly but the overall dynamics do not change over long timescales.}

We now describe our procedure for calculating \aveT\ as a function of $\Mp$ and $\Porb$, the results of which we present in \S~\ref{s:orbital decay}. 
For each reference network run at a given (\Mp, \Porb) point, we simulate at least $5\times10^5$ orbits in order to allow transient effects from initial conditions to die away.\footnote{\Revised{This is true of all networks in Figure \ref{f:multiparent} as well, with the exception of the largest $N_\mathrm{gens}=4$ network and the collective network, where we were computationally limited to shorter (but still convergent) integrations.}}
Using Equation (\ref{e:disp}),  we then compute the average $\dot{E}_\ast$  over the last $\sim 5\times10^4$ orbits of the integration.
We do this in order to average over the rapid fluctuations in dissipation that characterize the nonlinear equilibria, which correspond to r.m.s. variations at roughly a 10\% level.
We also find that $\dot{E}_\ast$  depends slightly (factor of $\approx 2$) on how close the parent is to a linear resonance peak (see the $N_{\rm gen}=3$ results in Figure \ref{f:detailed sweep}).  
In order to compute the average dissipation rate, we must therefore average the results over several resonance peaks. 
We do this by performing 21 separate integration runs, each at a slightly different orbital period ($\delta \Porb \ll \Porb$) chosen such that the runs span three resonance peaks.  
We then calculate the average $ \dot{E}_\ast$ weighted by the amount of time spent at that orbital period (Appendix \ref{s:time averaging}) and use this $\langle \dot{E}_\ast\rangle$ to estimate \aveT\ via Equation (\ref{e:tau}).  

\section{Results}\label{s:orbital decay}

\begin{figure}
        \begin{center}
    	\includegraphics[scale=0.73, clip=True, trim=0.0cm 0.4cm 1.0cm 0.50cm]{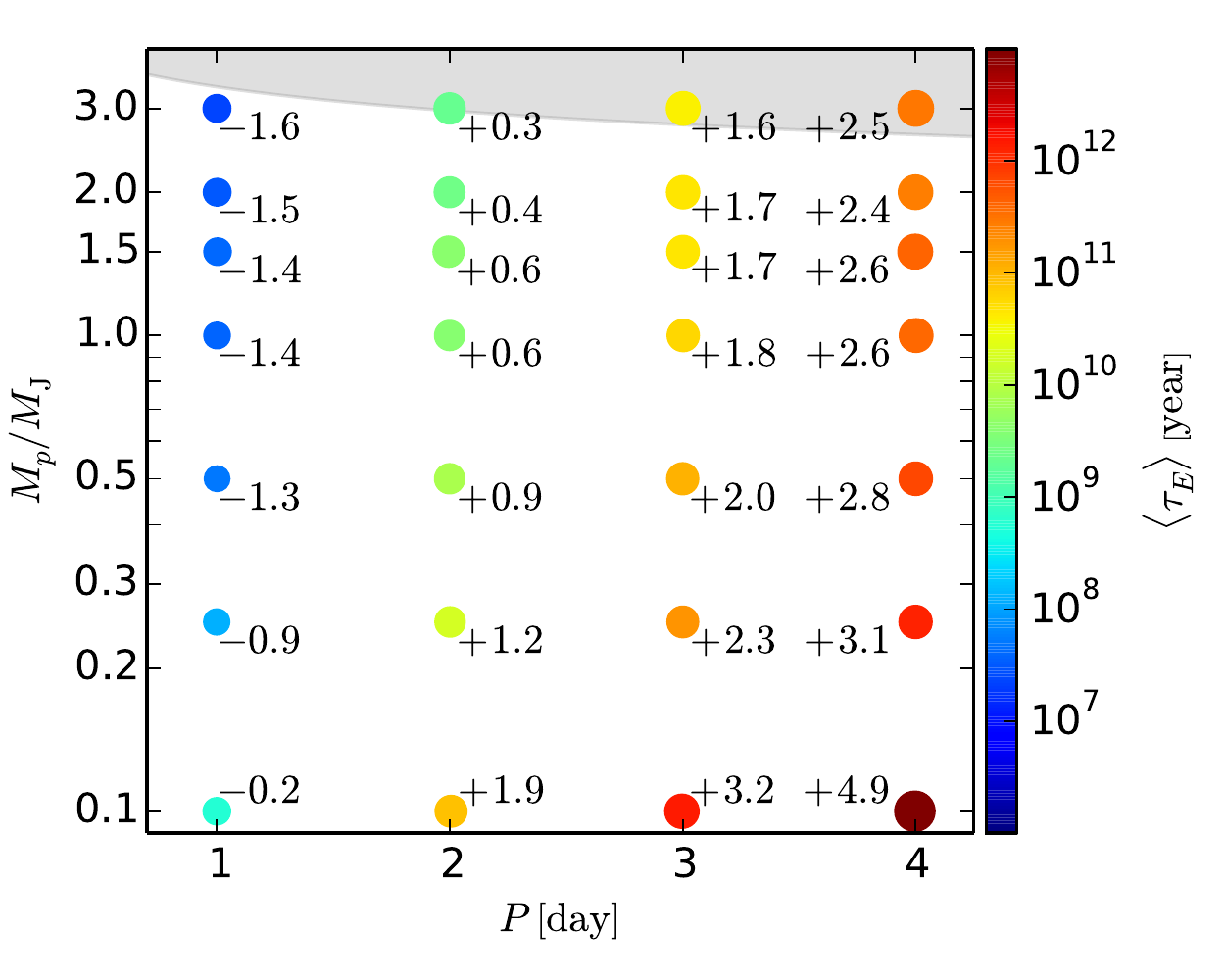}
        \end{center}
		\caption{
                Decay time \aveT\ on the ($\Mp$, $\Porb$) plane.
                The size of the marker is proportional to $\log_{10} \left( \Q \right)$, with corresponding color bar for \aveT\ in years.
                Each sample point is labeled with $\log_{10}\left(\aveT / {\rm Gyr}\right)$.
                The shaded region represents \cite{Barker:2011a}'s prediction for when linearly resonant modes break.
        }
        \label{f:mass-period plane}
\end{figure}

\begin{figure}
	\includegraphics[width=1.0\columnwidth, clip=True, trim=0.0cm 0.0cm 0.0cm 0.25cm]{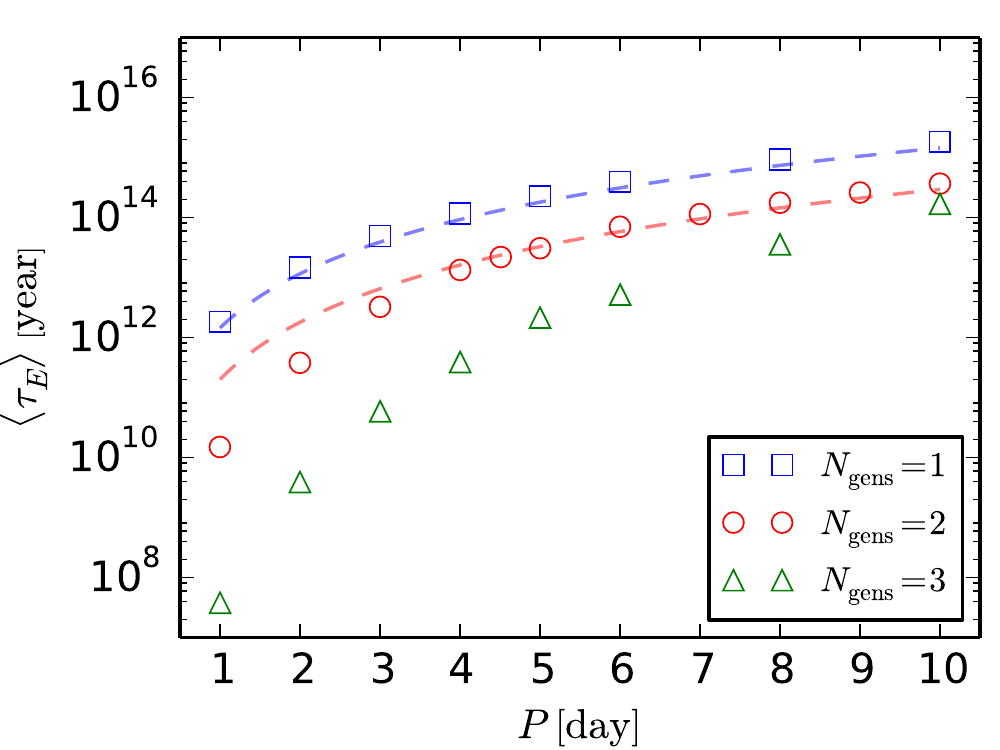}
	\caption{
		Decay time \aveT\ as a function of $\Porb$ for different numbers of generations and $\Mp=\Mj$.
        (blue squares) 10 most resonant parents.
        (red circles) parents+daughters.
        (green triangles) parents+daughters+granddaughters.
		(dashed lines) analytic estimates for $N_{\rm gens}=1,\, 2$ (Appendix \ref{s:Tlin}).
	}
	\label{f:frequency sweeps}
\end{figure}

In this section we present the results of integrating the coupled amplitude Equation (\ref{e:amp_eqn}) using the procedure and reference networks described in \S~\ref{s:saturation summation}. 
Figures \ref{f:mass-period decomposition} and \ref{f:mass-period plane} show \aveT\ as a function of $\Mp$ and $\Porb$.
We find that \aveT\ depends strongly on $\Porb$ and only mildly on $\Mp$. Our numerical results are well approximated by the fit 
\begin{equation}\label{e:tauE_final}
   \aveT = 4.1\times10^7 \left(\frac{\Mp}{\Mj}\right)^{-0.5} \left(\frac{\Porb}{\mathrm{day}}\right)^{6.7}\trm{ yr}
\end{equation}
\noindent
over the range $0.5\la \Mp/\Mj\le 3$ and $\Porb\la 4\trm{ days}$. This matches our numerical results to within a factor of $\simeq 2$ over this range, which is comparable to systematic modeling uncertainties due to differences in the network structure of large $N_{\rm gens}=3,4$ networks (see Figure \ref{f:multiparent}).  By Equation (\ref{e:Q}), 
this corresponds to a stellar tidal quality factor 
\begin{equation}\label{e:Qstar_final}
   \Q = 3.0\times10^5\left(\frac{\Mp}{\Mj}\right)^{0.5} \left(\frac{\Porb}{\mathrm{day}}\right)^{2.4}.
\end{equation}
\noindent
Thus, for $\Mp\ga 0.5 \Mj$ and $\Porb\la 2\trm{ day}$ we find that \aveT\ is small compared to the main-sequence lifetime of a solar-type star.  
For $\Mp \la 0.5\Mj$, we find that although \aveT\ increases significantly, it can still be small at small $\Porb$. 
For example, for $\Mp=0.1 \Mj$ and $\Porb=1\trm{ day}$ we find $\aveT \simeq 800 \trm{ Myr}$.
At sufficiently small $M_p$ and/or large $P$, the non-linear effects ``turn off'' and \aveT\ collapses onto the linear result. 
This is particularly evident for $\Mp=0.1\Mj$ in Figure \ref{f:mass-period decomposition}.
We also illustrate this effect in Figure \ref{f:frequency sweeps}, which shows how \aveT\ depends on \Porb\ when $\Mp=\Mj$ for different numbers of generations. 
We see that by $P\simeq10\trm{ days}$, \aveT\ is very long and close to the linear prediction (the $N_{\rm gen}=1$ line).
\begin{figure}
        \includegraphics[width=0.95\linewidth, clip=True, trim=0.0cm 0.55cm 0.05cm 0.50cm]{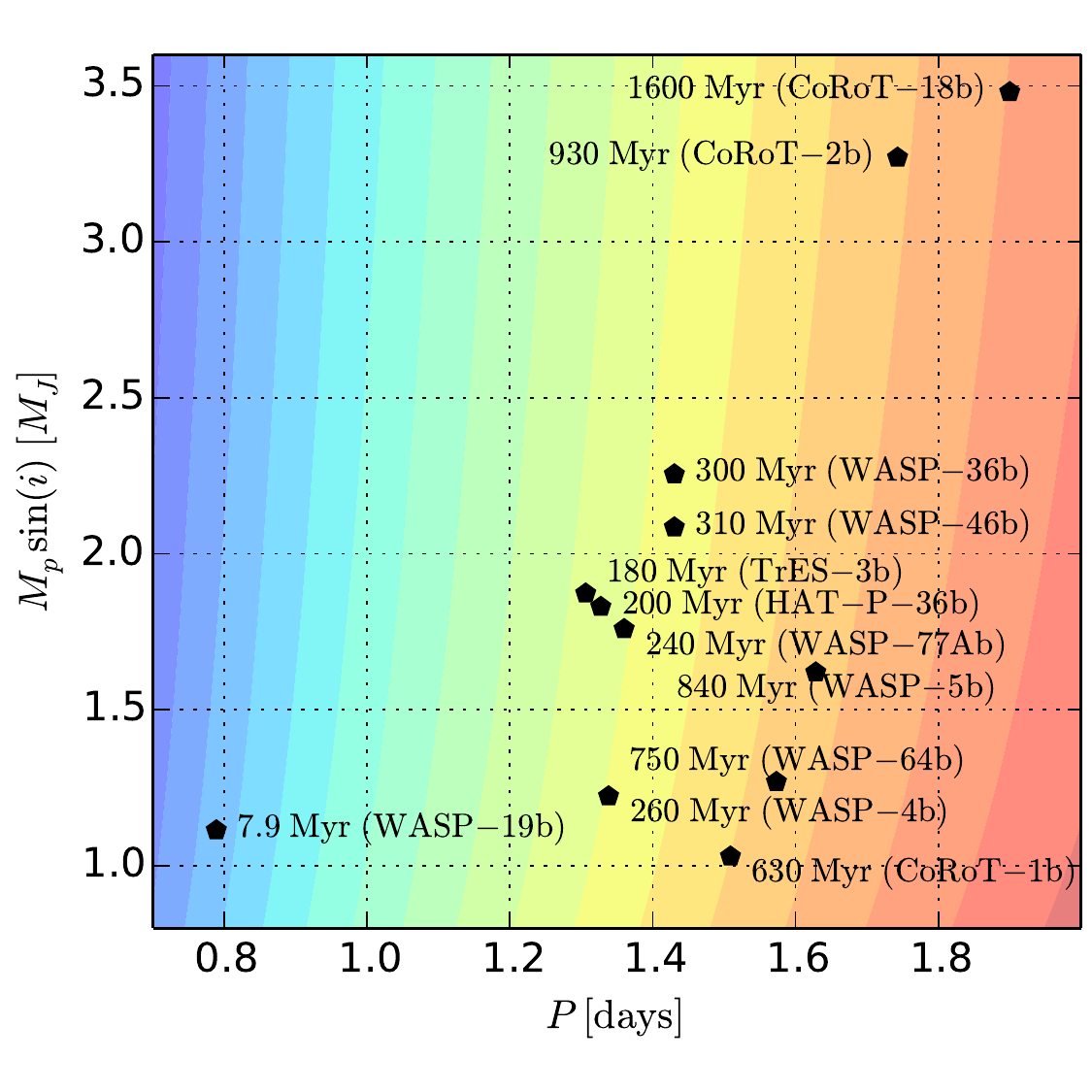}
        \caption{
                Period-mass distribution ($\Mp \sin i$ versus orbital period $\Porb$) of known extrasolar planets orbiting solar-type stars with $\aveT \lesssim 1\trm{ Gyr}$ based on our results.
                Each planet is labeled by our $\aveT$ fitting formula (Equation \ref{e:tauE_final}) , and the color represents contours of \aveT.
                (Data taken from  http://www.exoplanets.org).
                }
        \label{f:obs decay data}
\end{figure}
\begin{table*}
	\caption{
		Orbital parameters for known systems and a summary of predictions for the orbital decay timescales and change in orbital parameters after 10 years
	}
	\begin{center}
		\begin{tabular}{c||c|c|c|c}
    		                                            & WASP-19b                     & HAT-P-36b                    & WASP-36b                      & CoRoT-2b \\
		\hline
		\hline
		$M [M_\odot]$                               & $0.930\pm0.02$               & $1.022\pm 0.049$             & $1.020\pm0.032$               & $0.970 \pm 0.06$              \\
		$R [R_\odot]$                               & $0.990\pm0.02$               & $1.096 \pm 0.056$            & $0.943\pm0.019$               & $0.902 \pm 0.018$             \\
		$\Mp \sin i\ [\Mj]$     & $1.114\pm0.039$              & $1.83 \pm 0.1$               & $2.255\pm0.089$               & $3.27 \pm 0.171$              \\
   		$\Porb [\text{sec}]$                            & $68155.776\pm0.026$          & $114682.78\pm0.26$           & $132828.36\pm0.23$            & $150594.64\pm0.86$            \\
		eccentricity                                & $0.0046^{+0.0044}_{-0.0028}$ & $0.063\pm0.032$              & $0$                           & $0.0143^{+0.0077}_{-0.0076}$  \\
		age[\text{Gyr}]  &$10.2^{+3.0}_{-3.8}$ &   null     &  null  &$2.7^{+3.2}_{-2.7}$\\
		\hline
		$\aveT\equiv E_\mathrm{orb}/\left<\partial_t E_\ast\right> \ [\text{Myr}]$
                                                    & $9.2 \pm 0.128$              & $205 \pm 3.9$                & $454 \pm 15.7$                & $623 \pm 27.6$                \\
		$\text{min}\ \T [\text{Myr}]$               & $6.3$                        & $84.7$                       & $214$                         & $241$                         \\
		$\text{max}\ \T [\text{Myr}]$               & $12.4$                       & $311.$                       & $853$                         & $1150$                        \\
		\hline
		$\Delta \Porb \equiv (3 \Porb/ 2\T) \Delta t\ [\text{ms}]$           
		                                            & $110. \pm 1.5$               & $8.39 \pm 0.16$              & $4.4 \pm 0.15$                & $3.6 \pm 0.16$                \\
		$\text{min}\ \Delta \Porb [\text{ms}]$ 
                                                    & $82$                         & $5.5$                        & $2.3$                         & $1.96$                        \\
		$\text{max}\ \Delta \Porb [\text{ms}]$ 
		                                            & $161.$                       & $20.$                        & $9.3$                         & $9.38$                        \\
		\hline
		$T_\mathrm{shift} \equiv (3/4\T)(\Delta t)^2 \ [\text{sec}]$                   
		                                            & $257 \pm 3.6$                & $11.5 \pm 0.22$              & $5.2 \pm 0.18$                & $3.80 \pm 0.17$               \\
		$\text{min}\ T_\mathrm{shift}\ [\text{sec}]$
                                                    & $191$                        & $7.6$                        & $2.8$                         & $2.1$                         \\
		$\text{max}\ T_\mathrm{shift}\ [\text{sec}]$       
                                                    & $375$                        & $27.9$                       & $11.$                         & $9.8$                         \\
		\end{tabular}
	\end{center}
        \label{t:exoplanet predictions}
\end{table*}
\begin{figure*}
	\includegraphics[width=0.5\linewidth,, clip=True, trim=0.2cm -0.1cm 0.2cm 0.0cm]{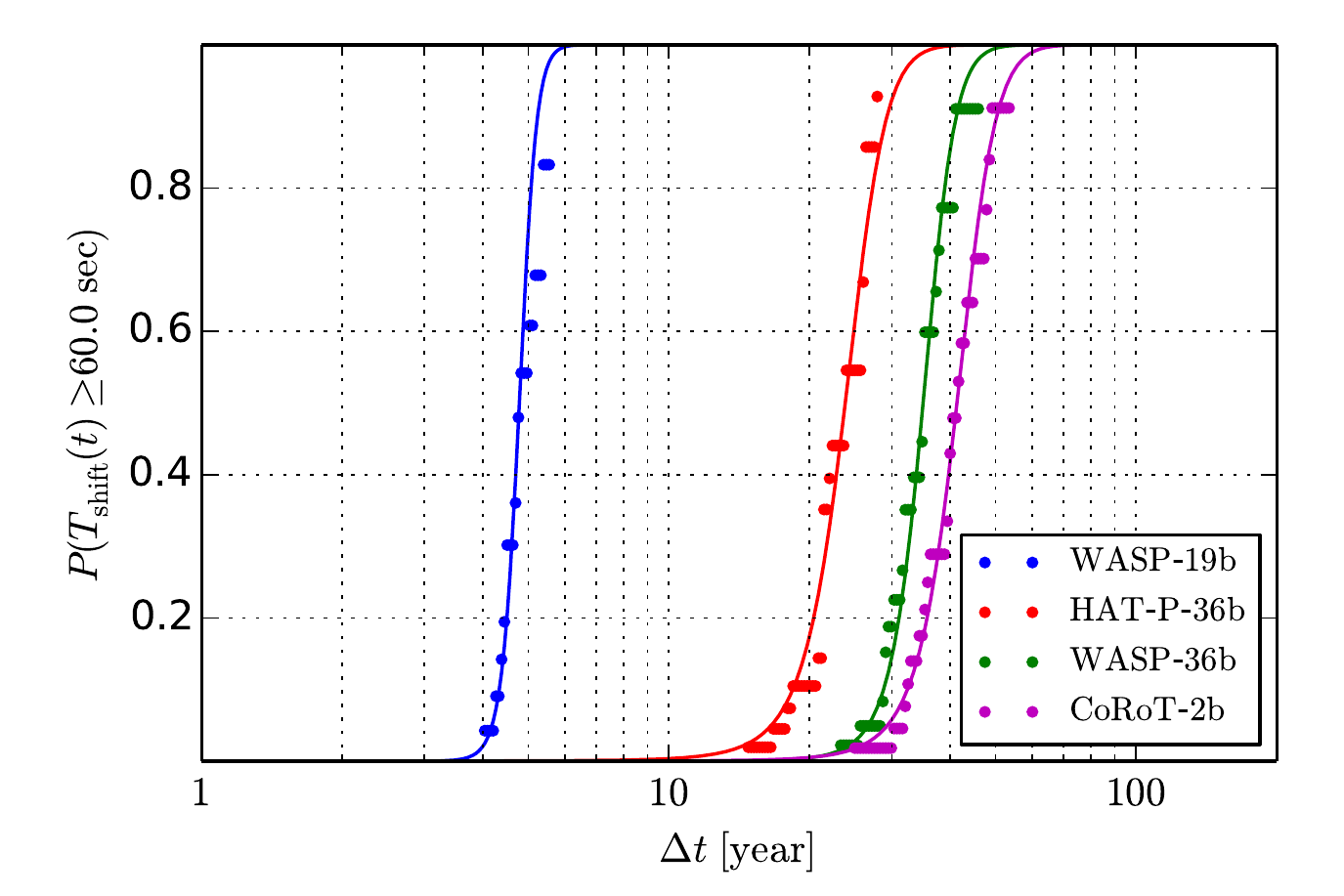} 
	\includegraphics[width=0.5\linewidth,, clip=True, trim=0.2cm -0.1cm 0.2cm 0.0cm]{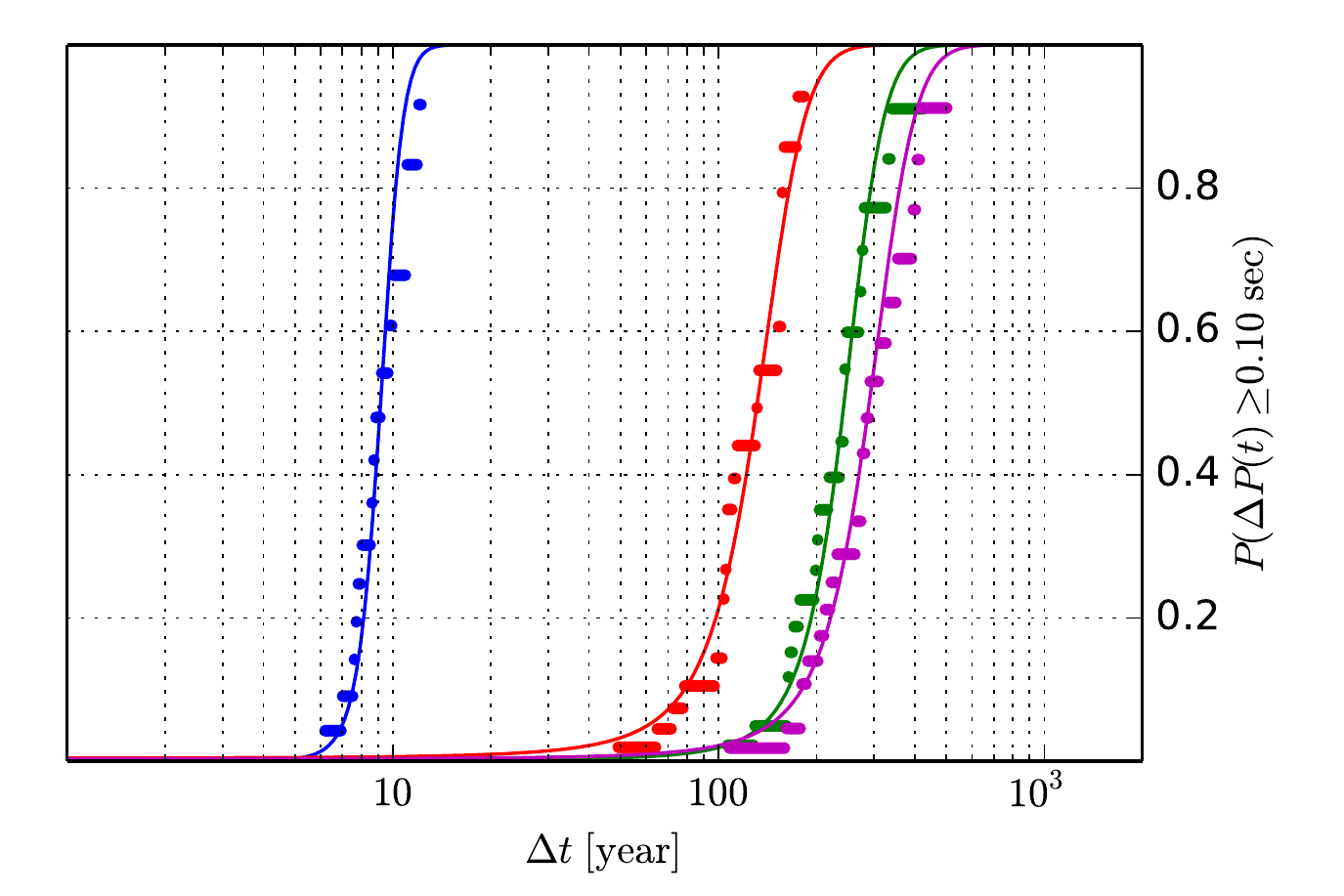}
	\caption{
		Cumulative probability distributions for observable changes as a function of elapsed time. 
                (blue) WASP-19b.
                (red) HAT-P-36b.
                (green) WASP-36b.
                (purple) CoRoT-2b.
        Solid lines represent approximate fits to the simulation results, which are represented by filled circles.}
	\label{f:cdf}
	\vspace{0.75cm}
\end{figure*}
%

\subsection{Implications for a few known systems}\label{s:known systems}
Based on the Exoplanet Orbit Database (http://www.exoplanets.org), there are currently 11 known planets orbiting approximately solar-type stars ($M=1.0\pm0.1 M_\odot$ and $T_{\rm eff}\simeq 5500\trm{ K}$) with decay times  $\aveT < 1\trm{ Gyr}$ according to our results.  
Of these, 7 have expected decay times $\aveT \la 0.3\trm{ Gyr}$; in order of increasing $\Porb$, they are WASP-19b, TrES-3b, HAT-P-36b, WASP-77Ab, WASP-4b, WASP-36b, and WASP-46b.
Figure \ref{f:obs decay data} shows these planets on the $\Mp\sin i$--$\Porb$ plane, with \aveT\ labeled for each system and a contour of \aveT\ superimposed.
These planets all have $\Mp \sin i > \Mj$, $\Porb<2.0\trm{ days}$, and eccentricities consistent with or very close to zero. \Revised{Since these are all transiting systems, $M_p \sin i \simeq M_p$ and the reported errors in the measured mass are typically $\la 0.1 \Mj$.}

We note that two of the planets (CoRoT-2b and CoRoT-18b) have masses $\Mp \sin i > 3\Mj$. 
This suggests that they are in the strongly nonlinear regime where the parent wave breaks within the stellar core \citep{Barker:2010, Barker:2011a, Barker:2011b}.
We discuss how our results compare to the strongly nonlinear simulations of Barker \& Ogilvie in \S~\ref{s:conclusions}.

\Revised{Of the 11 planets with  $\aveT < 1\trm{ Gyr}$, there are five for which studies report at least some constraint on the age of the system.  
In three of these, the age uncertainties are sufficiently large that the systems might be relatively young, i.e., $\sim 1\trm{ Gyr}$ (WASP-64b, WASP-5b, CoRoT-2b: $1.2^{+1.2}_{-0.7}$, $5.4^{+4.4}_{-4.3}$, $2.7^{+3.2}_{-2.7}$, respectively).  
However, WASP-4b and WASP-19b appear to be older systems: $7.0\pm 2.9$ and $10.2^{+3.0}_{-3.8}\trm {Gyr}$, respectively. 
Assuming that the planets arrived close to their current orbits when their host stars first formed,  such old stellar ages seem to be in tension with the small $\aveT$ we predict, especially in the case of WASP-19b.
If our results are correct, then perhaps these planets were scattered into their current orbits well after the stars formed or they just happened to initially reside at separations with decay timescales very close to their current ages.}

Several recent papers consider the prospects for the direct detection of orbital decay of individual planets by measuring transit timing variations (TTVs) over long time baselines ($\Delta t\ga 5\trm{ year}$, see \Revised{\citealt{Gandolfi:2014,Birkby:2014,Valsecchi:2014a,Watson:2010}}). 
In order to evaluate this possibility, we simulated four known systems spanning a variety of companion masses and orbital periods (but each with a solar-type host\footnote{This requirement is why we do not consider WASP-18b, which was analyzed in \cite{Birkby:2014}.}), calculating their tidally induced TTV ($T_{\rm shift}$) and change in orbital period ($\Delta \Porb$) as a function of $\Delta t$. 
We compute these according to (see \citealt{Birkby:2014} for a derivation)
\begin{subequations}
	\begin{align}
			T_\mathrm{shift}      & \approx \frac{1}{2} \frac{\dot{\Omega}}{\Omega} \left(\Delta t\right)^2 = \frac{3}{4\T}\left(\Delta t\right)^2,\\
				\Delta \Porb & \approx \dot{\Porb}\Delta t = \frac{3}{2} \frac{\Porb E_\mathrm{orb}}{\dot{E}_\mathrm{orb}} \Delta t = \frac{3 \Porb}{2\T} \Delta t.
	\end{align}
	\label{e:deviations}
\end{subequations}
\noindent
In order of increasing orbital period, we analyze WASP-19b \citep{Hebb:2010,Hellier:2011, Mortier:2013}, HAT-P-36b \citep{Bakos:2012}, WASP-36b \citep{Smith:2012}, and CoRoT-2b \citep{Alonso:2008,Gillon:2010}.

In order to calculate the orbital decay rate of these systems, we simulate a small range of orbital periods centered on each system's measured period.  We then compute the time-averaged decay rates following the procedure described in \S~\ref{s:saturation summation}.
We do this in order to mitigate any differences between the resonances of our stellar models and the actual resonances of the stellar hosts.
Furthermore, this allows us to compute a minimum and maximum expected decay rate, corresponding to the troughs and peaks of the resonances, respectively.

Table \ref{t:exoplanet predictions} lists $\aveT$ as well as the minimum and maximum $\T$. 
The $\aveT$ of the four systems ranges from about 10 Myr (WASP-19b) to 600 Myr (CoRoT-2b), while the minimum (maximum) $\T$ is approximately two times smaller (larger).
WASP-19b has by far the shortest decay time owing to its extremely short orbital period ($18.9$ hours).

Table \ref{t:exoplanet predictions} also lists the systems' average, minimum, and maximum $\Delta P$ and $T_{\rm shift}$ after ten years of evolution.
These provide an estimate of the magnitude of the tidally induced deviations we would expect to observe from these systems over the next ten years.

We quantify these effects further in Figure \ref{f:cdf}, which shows the cumulative probability of observing tide-induced deviations as a function of time.
We choose a detection threshold of $\left(\Delta \Porb\right)_\mathrm{thr}=0.1\ \mathrm{sec}$ and $\left(T_\mathrm{shift}\right)_\mathrm{thr}=60\ \mathrm{sec}$ based on the measurement errors of $\Porb$ and the expected uncertainties in TTVs \citep{Gillon:2009,Watson:2010}; different choices will scale $\Delta t$ through Equation (\ref{e:deviations}).

We find that $T_\mathrm{shift}$ should always produce a detection faster than $\Delta \Porb$. 
This is because $T_\mathrm{shift}$ is a cumulative effect that builds up throughout the orbital decay.  
According to our results, WASP-19b should produce a detectable $T_\mathrm{shift}$ in the very near future, with a $\approx50\%$ chance of observing a deviation now given the current four year baseline \citep{Hellier:2011} and a high likelihood of detection after only two more years.
It will take considerably longer before detections are possible in the other three systems.

We note that even if, for some reason, our calculations overestimate the dissipation rate by an order of magnitude, the $T_\mathrm{shift}$ curves in Figure \ref{f:cdf} would only be shifted to the right by a factor of $\sqrt{10}\sim 3$. Finally, as \cite{Watson:2010} point out, the Applegate effect could produce $\Delta \Porb$ and $T_\mathrm{shift}$ values that are comparable to the tidally induced values and distinguishing the two may not be simple. 

\subsection{Comparison with previous estimates of nonlinear tidal dissipation}\label{e:comparison_to_KG_BO}

Previous studies that attempt to estimate the nonlinear dissipation rate of dynamical tides in close binaries include  \citet{Kumar:1996} and \citet{Barker:2011a}. 
They both argue that an upper bound to the dissipation rate is approximately given by the product of the parent's linear energy and the three-mode growth rate of the fastest growing daughter pair: 
\beq
\dot{E}_\ast \la \Gamma_{\rm 3md} E_{\rm lin}.
\label{e:Edot_BO_estimate}
\eeq
This estimate does not account for the continuous linear driving of the parent. Instead, the parent wave is initialized with an energy equal to $E_{\rm lin}$ but is otherwise undriven, and the problem reduces to determining the amount of time it takes for daughters to dissipate that initial energy.  
Although this is appropriate for the tidal capture problem that \citet{Kumar:1996} consider (because the binary is on a highly eccentric orbit and the parent is only driven strongly during the brief pericenter passage), in our analysis the orbit is circular and the parent is a continuously driven standing wave.    
The estimate of Equation (\ref{e:Edot_BO_estimate}) also assumes that the mode dynamics are dominated by the single, fastest growing daughter pair even though there may be many modes participating in the interactions.

By Equation (\ref{e:s3}), we find that the fastest growing daughters have a growth rate
\begin{equation}
\Gamma_{\rm 3md}\simeq 0.6\left(\frac{M_p}{\Mj}\right)\left(\frac{P}{\trm{day}}\right)^{-11/6}\trm{ yr}^{-1}
\end{equation}
and by Equation (\ref{e:Elin})
\begin{equation}
 \Gamma_{\rm 3md} E_{\rm lin}\simeq 2.4\times10^{28}\left(\frac{M_p}{\Mj}\right)^3\left(\frac{P}{\trm{day}}\right)^{-7.5}\trm{ erg s}^{-1}.
\end{equation}
For comparison, the fit to our numerical simulations yields, by Equations (\ref{e:tauE_final}) and (\ref{e:instant tau}), 
\begin{equation}
\dot{E}_\ast\simeq 3.5\times10^{29}\left(\frac{M_p}{\Mj}\right)^{1.5}\left(\frac{P}{\trm{day}}\right)^{-7.4}\trm{ erg s}^{-1}.
\end{equation}
Thus, while the two have nearly identical $P$ scalings, the $\dot{E}_\ast$ from our simulations is $\simeq 15$ times larger than  $\Gamma_{\rm 3md} E_{\rm lin}$ for $M_p\simeq\Mj$.  This factor of 15 difference can be seen in the $\Q$ estimates.  In particular, we find $\Q \approx 3\times 10^5$ at $P=1\trm{ day}$ for $M_p \ga 0.5 \Mj$.  By contrast, \citet{Barker:2010} argue that $\Q \ga 5\times 10^6$ for systems below the wave breaking limit ($M_p \la 3\Mj$) based on their assumption that $\dot{E}_\ast \la \Gamma_{\rm 3md} E_{\rm lin}$ in the weakly nonlinear regime.

\begin{figure}
	\begin{center}
		\includegraphics[width=1.0\columnwidth, clip=True, trim=0.1cm 0.25cm 0.00cm 0.25cm]{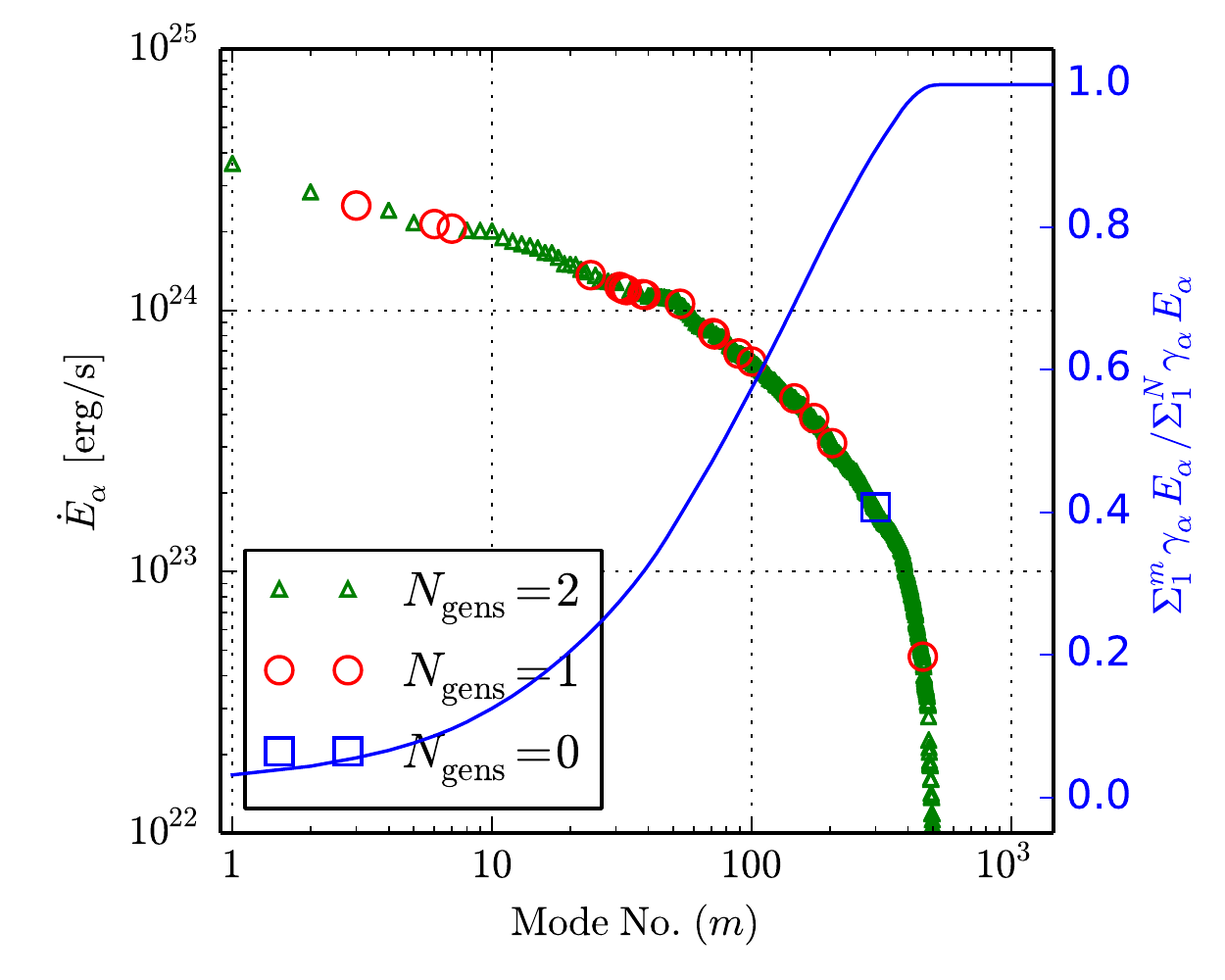}
	\end{center}
	\caption{
		Energy dissipation rate $\dot{E}_\alpha$ for each mode in our reference network, which consists of one parent (blue square), $\simeq 20$ daughters (red circles) and $\simeq 1500$ granddaughters (green triangles).  
		Modes are ordered by \Revised{$\dot{E}_{\alpha}$} and here we take $\Mp=\Mj$ and $\Porb\simeq 3\ {\rm days}$. 
		The blue line is the cumulative distribution of $\dot{E}_\alpha$; \Revised{we see} that granddaughter modes are responsible for the majority of the dissipation.
		}
	\label{f:yE_scatter}
\end{figure}

We suspect that the discrepancy is largely due to the assumption in Equation (\ref{e:Edot_BO_estimate}) that only the single fastest growing daughter pair is important.  In Figure \ref{f:yE_scatter} we demonstrate that this is not the case.  
We show the individual and cumulative contribution to $\dot{E}_\ast$ of modes in our reference network (which consists of $\simeq 1500$ modes).  We find that there are several daughter modes that contribute substantial amounts of dissipation, not just a single dominant daughter pair.
Figure \ref{f:yE_scatter} also shows that, in sum, the granddaughters are the dominant source of dissipation in the network.

\section{Summary and Discussion}\label{s:conclusions}

We present a first principles calculation of the saturation of nonlinear interactions between $g$-modes excited within the cores of solar-type hosts by planetary companions.
Using a WKB approximation for high-order, adiabatic $g$-modes and analytic approximations to their coupling coefficients detailed in WAQB, we systematically investigate the number of modes and types of couplings that are dynamically relevant. 
We determine the minimum mode network size and structure that yields total dissipation rates consistent with those of much larger networks (to within a factor of $\approx 2$).  
This minimum network is sufficiently nimble that we can efficiently explore broad swaths of the (\Mp, \Porb)-plane. We find that weakly nonlinear interactions are energetically important over large portions of this plane, including regions occupied by known exoplanetary systems.  
In these regions, the orbital decay time \aveT\ and stellar tidal quality factor \Q\ follow simple power law relations (Equations \ref{e:tauE_final} and \ref{e:Qstar_final}). 

 We find that the orbital decay of a number of observed hot Jupiters should occur on timescales much shorter than the main sequence lifetime of their host star.  
 Such rapid orbital decay could explain the observed paucity of giant planets with $P\la 2\trm{ days}$ (\citealt{McQuillan:2013}; see also \citealt{Winn:2014} for a recent review of the observations).  
 The short decay times would also induce TTVs that may be observable with current technology  (especially that of WASP-19b). 
 Precision photometry of individual systems may thus provide a new handle on tidal interactions within the next few years.

Our calculation comes with some caveats.  
First, although our reference network yields dissipation results that are very similar to those of the largest networks we investigate (which have $\ga 10$ times more modes than the reference network), there is still a possibility that the dynamics will change upon the addition of even more modes. 
Second, our calculation assumes that the modes are all global standing waves. 
However, this prescription may break down if the amplitudes of the modes change on timescales shorter  than the group travel times between their inner and outer turning points. 
Moreover, although the parent mode is below the wave breaking threshold (when not too close to a linear resonance), the daughter and granddaughter modes may not be.  
In Appendix \ref{s:breaking thr}, we show that the threshold amplitude of the three-mode parametric instability is much smaller than the wave breaking threshold and that both have the same frequency scaling. 
This may mean that further generations will be excited before the daughter and granddaughter modes break.  Nonetheless, this issue deserves further investigation.
Finally, we do not account for possible changes to the stellar structure due to the transfer of energy and angular momentum from the sea of excited waves.  
Further work is needed in order to determine the extent to which stellar spin-up, heating, and/or evolution affect background properties such as the star's stratification and thereby the wave interaction dynamics (see \citealt{Barker:2010} for a discussion of this issue).

Our study focuses on wave interactions in the weakly nonlinear regime.  
For solar type stars, this corresponds to planetary masses $M_p \la 3.6 \Mj (P / 1 \trm{ day})^{-0.1}$; above this mass, the parent wave breaks as it approaches the stellar center and the system is therefore in the strongly nonlinear regime \citep{Barker:2010, Barker:2011a, Barker:2011b}. 
In the weakly nonlinear regime the parent is a global standing wave while in the strongly nonlinear regime the parent is more appropriately treated as a traveling wave; it does not reflect upon reaching the stellar center. 
Barker \& Ogilvie study the fate of such a strongly nonlinear traveling wave with numerical simulations using a Boussinesq-type model. 
Because our calculation studies a different hydrodynamic regime, a direct comparison with their results is not possible.  
Nonetheless, one might expect the two to roughly agree near the region that marks the transition from weakly nonlinear to strongly nonlinear (i.e., near $M_p\simeq 3\Mj$). Indeed, \cite{Barker:2011b} finds $\Q \approx 9\times10^4(\Porb/\mathrm{day})^{2.8}$ for waves that break while we find $\Q \approx 5\times10^5(\Porb/\mathrm{day})^{2.4}$ for $M_p\simeq 3\Mj$. 
We explore some of the similarities between the two regimes further in Appendix \ref{s:breaking thr}.

We find $\Q \approx 3\times 10^5$ at $P=1\trm{ day}$ for $M_p \ga 0.5 \Mj$. 
This appears to conflict with the estimate in \citet{Barker:2010}, who argue that \Q\ increases rapidly to $\Q \ga 5\times 10^6$ for systems below the wave breaking limit ($M_p \la 3\Mj$).   
They do not attempt to calculate the saturation of the nonlinear parametric instabilities as in our study but instead base their estimate on stability analysis scaling arguments.  
As we explain in \S~\ref{e:comparison_to_KG_BO}, the issue might be that their estimate neglects the continuous driving of the parent and does not account for the complicated multi-mode dynamics that we find are important.  
Interestingly, we do see a steep increase in \aveT, although at much lower \Mp.

In order to be consistent with the observed distribution of exoplanets, \Revised{\cite{Penev:2012}} find that $\Q \ga 10^7$. 
However, as \cite{Birkby:2014} note, their analysis is for one specific set of initial conditions with some idealized assumptions about the chances of a planet candidate being confirmed by follow-up.   
They also assume gas disk migration and, as \Revised{\citeauthor{Penev:2012}} point out, their result may not be valid for other giant planet migration mechanisms such as dynamical scattering.  
If gas migration is the dominant mechanism that creates hot Jupiters, then our results suggest that finding these systems at $P\la 2\trm{ days}$ should be extremely rare.
However, if scattering populates short period orbits at random times after a system's formation, then a low \Q\ may not necessarily conflict with the observed population of hot Jupiters orbiting $\sim \trm{Gyr}$ old hosts.

Our study only considers solar-type hosts even though the observed population of hot Jupiters includes a wide variety of host types. Since the linear and nonlinear excitation of waves by the tide is sensitive to the detailed structure of the star, it is not clear how our results might depend on stellar type. Extending the analysis to non-solar type hosts would therefore allow us to more fully assess the prospects for measuring tide-induced orbital decay of individual hot Jupiter systems. 

\section{Acknowledgements}\label{s:acknowledgements}

\Revised{We thank Phil Arras and the referee for valuable comments on this manuscript.}
R.E. is supported in part by the National Science Foundation and
the LIGO Laboratory (PHY-0757058). 
This work was also supported by NASA NNX14AB40G. 


\appendix

\section{Computation of time-averages}\label{s:time averaging}

Quantities such as the energy dissipation rate $\dot{E}_\ast$ depend on how close the system happens to be to the densely spaced linear resonance peaks (the frequency spacing is $|\Delta_a|/\Omega\approx 10^{-2} (P/\trm{ day})^{-1}$). 
Because we are mostly interested in time-averaged statistics, at each ($M_p$, $\Porb$) point, we carry out 21 distinct simulations, each separated slightly in orbital period with a spacing chosen such that they span three resonance peaks (see \S~\ref{s:saturation summation} and Figure \ref{f:detailed sweep}). We compute the time-averaged statistic of a quantity $X$ by weighting each sample by the amount of time spent at that period 
\begin{equation}
	\left< X \right> = \frac{\int \mathrm{d}t X } {\int \mathrm{d} t} = \frac{\int dP \dot{P}^{-1} X } {\int \mathrm{d} \Porb \dot{P}^{-1}},
\end{equation}
\noindent 
where $\dot{P}=\mathrm{d}\Porb/\mathrm{d}t$ is the rate at which the period changes due to tidal dissipation.  
We compute $\dot{P}$  using an energy-balance argument. 
\Revised{We expect the time rate-of-change of the orbital energy ($E_\mathrm{orb}$), the tidal interaction-energy ($E_\mathrm{int}$), the rotational energy of the synchronized companion ($E_\mathrm{rot}$), and the energy stored in the modes ($E_\mathrm{modes}$) to balance with the energy lost through dissipation}
\begin{equation}
	\frac{\mathrm{d}}{\mathrm{d}t}\left( E_\mathrm{orb} + E_\mathrm{int} + E_\mathrm{rot} + E_\mathrm{modes} \right) = -2 \sum_i \gamma_i E_i
\end{equation}
\Revised{from which we can extract the time rate-of-change of the orbital period via}
\begin{eqnarray}
	\dot{P} & = & -2 \left(\frac{\mathrm{d} E_\mathrm{orb}}{\mathrm{d} P } + \frac{\mathrm{d} E_\mathrm{int}}{\mathrm{d} P } + \frac{\mathrm{d} E_\mathrm{rot}}{\mathrm{d} P} + \frac{\mathrm{d} E_\mathrm{modes}}{\mathrm{d} P }\right)^{-1} \sum_i \gamma_i E_i \\
	        & \approx & -2 \left(\frac{\mathrm{d} E_\mathrm{orb}}{\mathrm{d} P } \right)^{-1} \sum_i \gamma_i E_i \\
	        & \propto & \frac{P}{E_\mathrm{orb}} \sum_i \gamma_i E_i
\end{eqnarray}
\Revised{where we have noted that $|\mathrm{d}E_\mathrm{orb}/\mathrm{d}P| \gg |\mathrm{d}E_\mathrm{int}/\mathrm{d}P|,\ |\mathrm{d}E_\mathrm{rot}/\mathrm{d}P|,\ |\mathrm{d}E_\mathrm{modes}/\mathrm{d}P|$. 
This is because $E_\mathrm{orb}$ is much larger than any of the other energy scales, so even small relative changes in $E_\mathrm{orb}$ dominate over the other terms.}
This gives
\begin{equation}
                \left< X \right> = \frac{\int \mathrm{d} \Porb \left(\frac{\Porb}{E_\mathrm{orb}} \sum\limits_i \gamma_i E_i\right)^{-1} X }{\int \mathrm{d} \Porb \left(\frac{\Porb}{E_\mathrm{orb}} \sum\limits_i \gamma_i E_i\right)^{-1}} \approx \frac{ \sum\limits_\Porb{ \left(\frac{\Porb}{E_\mathrm{orb}} \sum\limits_i \gamma_i E_i\right)^{-1} X} } { \sum\limits_\Porb{ \left(\frac{\Porb}{E_\mathrm{orb}} \sum\limits_i \gamma_i E_i\right)^{-1}} }.
\end{equation}
We use this procedure to calculate the time-averaged $\dot{E}_\ast$ in the neighborhoods of each orbital period and \aveT.

\section{Three-mode non-linear equilibrium}\label{s:3mode equilib derivation}

Here we briefly review the non-linear equilibrium for three-mode systems.  The calculation is similar to that of Appendix D of WAQB except that here we provide more detail about the phase relations amongst the modes.  We begin with the equations of motion (Equation \ref{e:amp_eqn}) and introduce the change of coordinates $x = q e^{-i(\omega-\Delta)t}$, yielding
\begin{subequations}
	\begin{align}
                \partial_{t} x_{\alpha} + (i\Delta_{\alpha} + \gamma_{\alpha})x_{\alpha} & = i\omega_{\alpha} U_{\alpha} e^{-i (m_{\alpha}\Omega - \omega_\alpha + \Delta_\alpha) t} + 2i\omega_{\alpha} k_{\alpha\beta\gamma} x_{\beta}^\ast x_{\gamma}^\ast e^{i(\omega_\alpha + \omega_\beta + \omega_\gamma - \Delta_\alpha - \Delta_\beta - \Delta_\gamma)t} \\
                \partial_{t} x_{\beta} + (i\Delta_{\beta}+\gamma_{\beta})x_{\beta} & = 2i\omega_{\beta} k_{\alpha\beta\gamma} x_{\alpha}^\ast x_{\gamma}^\ast e^{i(\omega_\alpha + \omega_\beta + \omega_\gamma - \Delta_\alpha - \Delta_\beta - \Delta_\gamma)t} \\
                \partial_{t} x_{\gamma} + (i\Delta_{\gamma}+\gamma_{\gamma})x_{\gamma} & = 2i\omega_{\gamma} k_{\alpha\beta\gamma} x_{\alpha}^\ast x_{\beta}^\ast e^{i(\omega_\alpha + \omega_\beta + \omega_\gamma - \Delta_\alpha - \Delta_\beta - \Delta_\gamma)t}.
	\end{align}
\end{subequations}
We can cancel all time dependence in these equations by demanding
\begin{equation}
		m_\alpha\Omega = \omega_\alpha - \Delta_\alpha = \Delta_\beta + \Delta_\gamma - \omega_\beta - \omega_\gamma
\end{equation}
and assuming that $\partial_t \rightarrow 0$ in order to explicitly seek time independent solutions.
Manipulating the two daughter equations yields
\begin{equation}
		(i\Delta_\beta + \gamma_\beta) x_\beta (-i\Delta_\gamma + \gamma_\gamma ) = 2i\omega_\beta k_{\alpha\beta\gamma} x_\alpha^\ast \left( (i\Delta_\gamma + \gamma_\gamma) x_\gamma \right)^\ast = 4 \omega_\beta \omega_\gamma k_{\alpha\beta\gamma}^2 x_\alpha x_\alpha^\ast x_\beta
\end{equation}
which implies
\begin{equation}
	\Delta_\beta \gamma_\gamma = \Delta_\gamma \gamma_\beta, \hspace{1cm} \Delta_\beta \Delta_\gamma + \gamma_\beta \gamma_\gamma = 4 \omega_\beta \omega_\gamma k_{\alpha\beta\gamma}^2 A_\alpha^2 \label{e:delta-gamma},
\end{equation}
where we write $x=Ae^{i\delta}$. We then have
\begin{equation}
	A_\alpha^2 = \frac{\Delta_\beta \Delta_\gamma + \gamma_\beta \gamma_\gamma}{4 \omega_\beta \omega_\gamma k_{\alpha\beta\gamma}^2} = \frac{\gamma_\beta \gamma_\gamma}{4 \omega_\beta \omega_\gamma k_{\alpha\beta\gamma}^2} \left[ 1 + \left(\frac{ \Delta_\beta + \Delta_\gamma }{ \gamma_\beta + \gamma_\gamma}\right)^2 \right]
\end{equation}
and we recover the parent instability threshold energy $E_{\rm thr}=A_\alpha^2$ (Equation \ref{e:Ethr}). 
The daughter equations yield
\begin{equation}
	\frac{(i\Delta_\beta + \gamma_\beta)x_\beta}{\omega_\beta x_\gamma} = \frac{(i\Delta_\gamma + \gamma_\gamma)x_\gamma}{\omega_\gamma x_\beta} \Longrightarrow \left( \frac{A_\beta}{A_\gamma} \right)^2 = \frac{\gamma_\gamma \omega_\beta}{\gamma_\beta \omega_\gamma},
\end{equation}
which gives
\begin{equation}
	(i\Delta_\beta + \gamma_\beta) A_\beta = 2i\omega_\beta k_{\alpha\beta\gamma} A_\alpha A_\gamma e^{-i(\delta_\alpha + \delta_\beta + \delta_\gamma)}, \\
\end{equation}		
or equivalently
\begin{equation}
	\gamma_\beta \frac{A_\beta}{A_\gamma} = 2\omega_\beta k_{\alpha\beta\gamma} A_\alpha \sin \delta, \hspace{1cm} \Delta_\beta \frac{A_\beta}{A_\gamma} = 2\omega_\beta k_{\alpha\beta\gamma} A_\alpha \cos \delta,
\end{equation}
where $\delta = \delta_\alpha + \delta_\beta + \delta_\gamma$. 
We can now use the parent equation to determine the parent phase $\delta_\alpha$ and the product of the daughter amplitudes
\begin{equation}
	(i\Delta_\alpha + \gamma_\alpha)A_\alpha = i\omega_\alpha U_\alpha e^{-i\delta_\alpha} + 2 i \omega_\alpha k_{\alpha\beta\gamma} A_\beta A_\gamma e^{-i\delta}.
\end{equation}
After some manipulation, we find
\begin{equation}
	A_\beta A_\gamma = \left( \frac{A_\alpha}{2\omega_\alpha \kappa_{\alpha\beta\gamma}}\right) \left[ \left(\Delta_\alpha \cos\delta + \gamma_\alpha \sin\delta\right) \pm \sqrt{\left(\Delta_\alpha \cos\delta + \gamma_\alpha \sin\delta\right)^2 + \frac{\Delta_\alpha^2 + \gamma_\alpha^2}{A_\alpha^2} \left(\frac{\omega_\alpha^2 U_\alpha^2}{\Delta_\alpha^2 + \gamma_\alpha^2} - A_\alpha^2\right) } \right].
\end{equation}
The choice of sign depends on the sign of $\omega_\alpha\kappa_{\alpha\beta\gamma}$ and is determined by the requirement that the daughter amplitudes be positive.  Finally, by Equation (\ref{e:Elin}), we see that the instability condition is
\begin{equation}
	\frac{\omega_{\alpha}^2 U_{\alpha}^2}{\Delta_{\alpha}^2+\gamma_{\alpha}^2} = \frac{E_\mathrm{lin}}{E_0} > A_\alpha^2 = \frac{E_{\rm thr}}{E_0}.
\end{equation}
Note that we can solve for the parent's phase $\delta_\alpha$ and the sum of all the mode phases $\delta$, but we cannot break the degeneracy between the daughters' phases.  This is observed numerically, and carries information about the initial conditions.

\section{Two daughters, N parents}\label{s:2daughter Nparent}

If we linearize around the linear-equilibrium solution, the equations of motion for the daughter modes become
\begin{equation}
	\partial_t q_\beta + \left(i\omega_\beta + \gamma_\beta\right) = 2i\omega_\beta q_\gamma^\ast \sum\limits_{p\,\in\,{\rm  parents}} \kappa_{p\beta\gamma} q_p^\ast
\end{equation}
for daughter $\beta$ and the equivalent equation with the exchange $\gamma \leftrightarrow \beta$ for daughter $\gamma$. 
We can analyze this system as if there is a single parent with complex amplitude
\begin{equation}
        \kappa q = \sum\limits_{p\,\in\,{\rm  parents}} \kappa_{p\beta\gamma} q_p^\ast.
\end{equation}

We note the possibility for parent modes to interfere with one another when driving daughter modes, possibly rendering daughters stable under multi-parent driving when they were unstable to any individual parent.
Most notably, if the parents are nearly regularly spaced in frequency and driven at the midpoint between their resonance peaks, there can be strong destructive interference.
This is because each parent is paired with a partner on the opposite side of the driving frequency, and each pair consists of parents oscillating with nearly opposite phases.
This narrow ``trap'' in the resonance troughs is readily apparent at orbital periods above 4 days for a solar-type host of a Jupiter mass companion.
However, we did not observe significant ``trapping'' below $\sim$ 4 day orbital periods, where we focus our attention for this study.
This may be due to the asymmetric spacing of resonances, which will destroy this near perfect cancellation, or due to the amplitudes being large enough to overcome any cancellation that was present. In the hot Jupiter context, this issue is probably only of theoretical interest since the orbital evolution time scales are $\ga 10^{11}\trm{ yr}$ for $P \ga 3\trm{ day}$, even for a $3\, \Mj$ companion.

\section{details of collective set selection algorithm}\label{s:collective}

One can easily think of more complicated collective sets than what is described in \S~\ref{s:selecting collective sets}.
We analyze several of these systems in Appendix \ref{s:collective stability}.
In order to detect and include the diverse set of collective systems, we implement a broad search through parameter space.
We begin with a ``seed'' triple in parameter space, typically taken to be a minima of $E_\mathrm{thr}$.
We then expand the set of included modes in ($n$, $l$)-space around these seeds, choosing new modes from the border of the included set.
For these border modes, we compute the three-mode $E_\mathrm{thr}$ for all possible couplings between that border mode and the interior modes.
We then sort these $E_\mathrm{thr}$, and divide each by the number of couplings that produce $E_\mathrm{thr}$ less than or equal to the current value.
We take the minimum ratio and call it the ``collective $E_\mathrm{thr}$.''
This approximates the scaling with $N$ predicted in Appendix \ref{s:collective stability} and incorporates the decoupling of large detuning modes discussed in \S~\ref{s:decoupling of large detuning}.
Border modes are added in order of increasing collective $E_\mathrm{thr}$, and these thresholds are updated each time a mode is added to the network.
If the detuning increases the three-mode $E_\mathrm{thr}$ faster than the number of modes included, then small sets with low detuning will naturally be chosen.
However, if the detuning increases $E_\mathrm{thr}$ more slowly than the number of modes included, then the collective $E_\mathrm{thr}$ will decrease with the addition of more modes and the algorithm will select a set of collectively unstable daughters.

We typically find that a minimum number of daughters is needed before the scaling with $N$ dominates over the increase in $E_\mathrm{thr}$. 
Depending on $E_\mathrm{lin}$, these collective sets can grow to several thousand modes.
Although each mode can only directly couple to a relatively small number of other modes (see \S~\ref{s:selecting collective sets} and WAQB), we find that many smaller sets overlap and are thereby strung together to create larger networks. We discuss some of this behavior in \S~\ref{s:overlapping collective set}.

This algorithm scales poorly with the number of modes included ($O[N^3]$). Furthermore, as we describe in \S~\ref{s:mode dynamics}, we find that we can accurately model the total dissipation within the star using only three-mode pairs, rather than collective sets.
This, coupled with the fact that large collective networks are expensive to integrate, is the reason we choose three-mode networks with many couplings and generations as our reference networks discussed in \S\S~\ref{s:saturation summation}, \ref{s:orbital decay}, and \ref{s:known systems}.
We note, however, that collective sets may be important if one is interested in accurately modeling the dynamics of any particular mode, rather than the network as a whole. 

\section{Collective set stability}\label{s:collective stability}

In this appendix we analyze collective instabilities, i.e., sets of daughter modes that display rapid growth rates due to their mutual inter-coupling. Although in our simulations we find that they do not contribute significantly to the total tidal dissipation in hot Jupiter systems, for completeness we present here derivations of different stability thresholds for different types of collective instabilities.  Our mode selection algorithm (Appendix \ref{s:collective}) finds complicated collective sets that contain these types of coupling topologies.

\subsection{Single collective set}\label{s:single collective set}

We first consider the stability of a single collective set.  Since we are interested in the stability of linear solutions, we can assume that the parent is at a fixed amplitude
\begin{equation}
q_o = A_o e^{-i(\Omega t -\delta_o)}.
\end{equation}
The equation of motion of each daughter is then
\begin{eqnarray}
\dot{q}_\alpha + (i\omega_\alpha+\gamma_\alpha)q_\alpha & = & i\omega_\alpha \sum_{\beta} \kappa_{o\alpha\beta} A_o e^{+i(\Omega t -\delta_o)} q_\beta^\ast \\
& = & i\omega_\alpha\kappa_{o\alpha\alpha}A_o e^{+i(\Omega t -\delta_o)} q_\alpha^\ast + 2i\omega_\alpha\sum_{\beta\neq\alpha}\kappa_{o\alpha\beta}A_o e^{+i(\Omega t -\delta_o)} q_\beta^\ast. 
\end{eqnarray}
Defining a new set of variables $q = xe^{-i(\omega -\Delta)t}$, we can re-write the daughter equations as
\begin{eqnarray}
\dot{x}_\alpha + (i\Delta_\alpha+\gamma_\alpha)x_\alpha & = & i\omega_\alpha\kappa_{o\alpha\alpha}A_o x_\beta^\ast e^{+i(\Omega +2\omega_\alpha-2\Delta_\alpha)t-i\delta_o} \\
& &  + 2i\omega_\alpha\sum_{\beta\neq\alpha}\kappa_{o\alpha\beta}A_o x_\beta^\ast e^{+i(\Omega +\omega_\alpha+\omega_\beta -\Delta_\alpha-\Delta_\beta)t -i\delta_o} \\
& = & i\omega_\alpha\kappa_{o\alpha\alpha}A_o x_\beta^\ast e^{-i\delta_o} + 2i\omega_\alpha\sum_{\beta\neq\alpha}\kappa_{o\alpha\beta}A_o x_\beta^\ast e^{-i\delta_o}, 
\end{eqnarray}
where in the last step we demanded that the time dependence cancels
\begin{equation}
\Omega +\omega_\alpha+\omega_\beta -\Delta_\alpha-\Delta_\beta = 0\ \forall\ \{\alpha,\beta\}.
\end{equation}
Analyzing this as an eigenvalue problem, separate $x$ into real and imaginary parts $x=R+iI$,
\begin{equation}
\begin{bmatrix}
\dot{R}_\alpha \\
\dot{I}_\alpha 
\end{bmatrix}
=
\begin{bmatrix}
-\gamma_\alpha + \omega_\alpha A_o \kappa_{o\alpha\alpha}\sin\delta_o & \Delta_\alpha + \omega_\alpha A_o \kappa_{o\alpha\alpha}\cos\delta_o \\
-\Delta_\alpha + \omega_\alpha A_o \kappa_{o\alpha\alpha}\cos\delta_o & -\gamma_\alpha - \omega_\alpha A_o \kappa_{o\alpha\alpha}\sin\delta_o 
\end{bmatrix}
\begin{bmatrix}
R_\alpha \\
I_\alpha \\
\end{bmatrix}
+ \sum_{\beta\neq\alpha} \begin{bmatrix}
2\omega_\alpha A_o \kappa_{o\alpha\beta}\sin\delta_o & 2\omega_\alpha A_o \kappa_{o\alpha\beta}\cos\delta_o \\
2\omega_\alpha A_o \kappa_{o\alpha\beta}\cos\delta_o & -2\omega_\alpha A_o \kappa_{o\alpha\beta}\sin\delta_o 
\end{bmatrix}
\begin{bmatrix}
R_\beta \\
I_\beta \\
\end{bmatrix}.
\end{equation}
If we assume $[R_\alpha, I_\alpha] \propto e^{st}\ \forall\ \alpha$, then this equation becomes
\begin{multline}
0
=
\begin{bmatrix}
-(\gamma_\alpha+s) + \omega_\alpha A_o \kappa_{o\alpha\alpha}\sin\delta_o & \Delta_\alpha + \omega_\alpha A_o \kappa_{o\alpha\alpha}\cos\delta_o \\
-\Delta_\alpha + \omega_\alpha A_o \kappa_{o\alpha\alpha}\cos\delta_o & -(\gamma_\alpha+s) - \omega_\alpha A_o \kappa_{o\alpha\alpha}\sin\delta_o 
\end{bmatrix}
\begin{bmatrix}
R_\alpha \\
I_\alpha \\
\end{bmatrix}
\\ 
+ \sum_{\beta\neq\alpha} \begin{bmatrix}
2\omega_\alpha A_o \kappa_{o\alpha\beta}\sin\delta_o & 2\omega_\alpha A_o \kappa_{o\alpha\beta}\cos\delta_o \\
2\omega_\alpha A_o \kappa_{o\alpha\beta}\cos\delta_o & -2\omega_\alpha A_o \kappa_{o\alpha\beta}\sin\delta_o 
\end{bmatrix}
\begin{bmatrix}
R_\beta \\
I_\beta \\
\end{bmatrix}.
\end{multline}

This is an eigenvalue problem for a large matrix and the general decomposition is difficult. However, the matrix can be made almost symmetric and if we make several approximations the problem becomes analytically tractable. Specifically, if we assume
\begin{equation}\label{e:collective_assumptions}
\left.\begin{matrix}
\omega_\alpha = \omega \\ 
\gamma_\alpha = \gamma \\
\kappa_{o\alpha\alpha} = \kappa_s 
\end{matrix}\ \right|\ \forall\ \alpha
, \hspace{1cm}
\kappa_{o\alpha\beta} = \kappa\ \forall\ \alpha\neq\beta,
\end{equation}
\noindent
then we can define
\begin{equation}
M_S \equiv 
\begin{bmatrix}
-(\gamma+s) + \omega A_o \kappa_s\sin\delta_o & \Delta + \omega A_o \kappa_s\cos\delta_o \\
-\Delta + \omega A_o \kappa_s\cos\delta_o & -(\gamma+s) - \omega A_o \kappa_s\sin\delta_o
\end{bmatrix}
, \hspace{1cm}
M_I \equiv
\begin{bmatrix}
2\omega A_o \kappa\sin\delta_o & 2\omega A_o \kappa\cos\delta_o \\
2\omega A_o \kappa\cos\delta_o & -2\omega A_o \kappa\sin\delta_o
\end{bmatrix},
\end{equation}
\noindent
where $\Delta_\alpha = \Delta\ \forall\ \alpha$ since $\omega_\alpha = \omega\ \forall\ \alpha$. 
Writing this as a single matrix and requiring non-trivial mode amplitudes, we obtain
\begin{equation}
0=\mathrm{det}
\begin{vmatrix}
M_S & M_I & M_I & M_I & \cdots & M_I & M_I \\
M_I & M_S & M_I & M_I & \cdots & M_I & M_I \\
M_I & M_I & M_S & M_I & \cdots & M_I & M_I \\
M_I & M_I & M_I & M_S & \cdots & M_I & M_I \\
\vdots & \vdots & \vdots & \vdots & \ddots & \vdots & \vdots \\
M_I & M_I & M_I & M_I & \cdots & M_S & M_I \\
M_I & M_I & M_I & M_I & \cdots & M_I & M_S 
\end{vmatrix}
= \mathrm{det}\left|M_S-M_I\right|^{N-1}\mathrm{det}\left|M_S+(N-1)M_I\right|.
\end{equation}
We have $N-1$ repeated pairs of roots and one additional pair. The eigenvalues can be easily computed from
\begin{eqnarray}
\mathrm{det}\left|M_S-M_I\right| & = & (\gamma+s)^2 +\Delta^2 - \omega^2 A_o^2(\kappa_s-2\kappa)^2 = 0 \nonumber \\
\Rightarrow s & = & -\gamma \pm \sqrt{\omega^2 A_o^2 (2\kappa - \kappa_s)^2 - \Delta^2} 
\end{eqnarray}
and
\begin{eqnarray}
\mathrm{det}\left|M_S+(N-1)M_I\right| & = & (\gamma+s)^2 + \Delta^2 - \omega^2 A_o^2 (2(N-1)\kappa + \kappa_s)^2  =  0 \nonumber \\
\Rightarrow s & = & -\gamma \pm \sqrt{\omega^2 A_o^2 (2(N-1)\kappa + \kappa_s)^2 - \Delta^2}. 
\end{eqnarray}
In particular, we are interested in the values of $A_o$ for which $\mathbb{R}\{s\}\rightarrow 0$. These are
\begin{equation}
	A_{\rm thr}^2  = \frac{\gamma^2 + \Delta^2}{4\omega^2(\kappa-\frac{1}{2}\kappa_s)^2} = \frac{\gamma\gamma}{4\omega\omega(\kappa-\frac{1}{2}\kappa_s)^2}\left[1 + \frac{(\Delta+\Delta)^2}{(\gamma+\gamma)^2}\right]
\end{equation}
and
\begin{equation}
	A_{\rm thr}^2 = \frac{\gamma^2 + \Delta^2}{4\omega^2((N-1)\kappa + \frac{1}{2}\kappa_s)^2} = \left(\frac{1}{N-1}\right)^2\frac{\gamma\gamma}{4\omega\omega(\kappa + \frac{1}{2(N-1)}\kappa_s)^2}\left[ 1 + \frac{(\Delta+\Delta)^2}{(\gamma+\gamma)^2}\right],
\end{equation}
respectively. 
We see that there are $N-1$ modes that resemble ``standard'' three-mode instabilities and one collective eigenvalue, with an amplitude threshold suppressed by a factor of $N-1$.

Because of the assumptions in Equation (\ref{e:collective_assumptions}), the actual value of $A_{\rm thr}$ will differ somewhat from this expression. Nonetheless, we expect it to generalize to the requirement that 
\begin{equation}
(N-1)^2 A_o^2 \ga \frac{\gamma_1 \gamma_2}{4\omega_1\omega_2\kappa_{o12}^2}\left[ 1 + \frac{(\Delta_1+\Delta_2)^2}{(\gamma_1+\gamma_2)^2}\right] \ \forall\ \mathrm{modes} \ 1,2 \in \mathrm{collective\ set\ of}\ N \ \mathrm{modes},
\end{equation}
where $\Delta_1+\Delta_2 = \Omega+\omega_1+\omega_2$.

\subsection{Overlapping collective modes stability}\label{s:overlapping collective set}

We now consider a coupling topology where there are three types of modes. 
The $A$ modes are coupled to other $A$ modes and to $C$ modes.
$B$ modes are coupled to other $B$ modes and to $C$ modes.
$C$ modes are coupled to all other modes. Furthermore, we assume that all $A$, $B$, and $C$ modes are coupled to the same parent modes, which we treat as a single parent even though multiple parents may be acting (see Appendix \ref{s:2daughter Nparent}).

The associated eigenvalue problem yields the following characteristic equation
\begin{equation}
0=\mathrm{det}
\begin{vmatrix}
M_S^A  & M_I^A  & \cdots & M_I^A  & M_I^A  & M_I^A  & M_I^A  & \cdots & M_I^A  & M_I^A  & 0      & 0      & \cdots & 0      & 0      \\
M_I^A  & M_S^A  & \cdots & M_I^A  & M_I^A  & M_I^A  & M_I^A  & \cdots & M_I^A  & M_I^A  & 0      & 0      & \cdots & 0      & 0      \\
\vdots & \vdots &        & \vdots & \vdots & \vdots & \vdots &        & \vdots & \vdots & \vdots & \vdots & \vdots & \vdots & \vdots \\ 
M_I^A  & M_I^A  & \cdots & M_S^A  & M_I^A  & M_I^A  & M_I^A  & \cdots & M_I^A  & M_I^A  & 0      & 0      & \cdots & 0      & 0      \\
M_I^A  & M_I^A  & \cdots & M_I^A  & M_S^A  & M_I^A  & M_I^A  & \cdots & M_I^A  & M_I^A  & 0      & 0      & \cdots & 0      & 0      \\
M_I^A  & M_I^A  & \cdots & M_I^A  & M_I^A  & M_S^C  & M_I^C  & \cdots & M_I^C  & M_I^C  & M_I^B  & M_I^B  & \cdots & M_I^B  & M_I^B  \\
M_I^A  & M_I^A  & \cdots & M_I^A  & M_I^A  & M_I^C  & M_S^C  & \cdots & M_I^C  & M_I^C  & M_I^B  & M_I^B  & \cdots & M_I^B  & M_I^B  \\
\vdots & \vdots &        & \vdots & \vdots & \vdots & \vdots &        & \vdots & \vdots & \vdots & \vdots &        & \vdots & \vdots \\
M_I^A  & M_I^A  & \cdots & M_I^A  & M_I^A  & M_I^C  & M_I^C  & \cdots & M_S^C  & M_I^C  & M_I^B  & M_I^B  & \cdots & M_I^B  & M_I^B  \\
M_I^A  & M_I^A  & \cdots & M_I^A  & M_I^A  & M_I^C  & M_I^C  & \cdots & M_I^C  & M_S^C  & M_I^B  & M_I^B  & \cdots & M_I^B  & M_I^B  \\
0      & 0      & \cdots & 0      & 0      & M_I^B  & M_I^B  & \cdots & M_I^B  & M_I^B  & M_S^B  & M_I^B  & \cdots & M_I^B  & M_I^B  \\
0      & 0      & \cdots & 0      & 0      & M_I^B  & M_I^B  & \cdots & M_I^B  & M_I^B  & M_I^B  & M_S^B  & \cdots & M_I^B  & M_I^B  \\
\vdots & \vdots &        & \vdots & \vdots & \vdots & \vdots &        & \vdots & \vdots & \vdots & \vdots &        & \vdots & \vdots \\
0      & 0      & \cdots & 0      & 0      & M_I^B  & M_I^B  & \cdots & M_I^B  & M_I^B  & M_I^B  & M_I^B  & \cdots & M_S^B  & M_I^B  \\
0      & 0      & \cdots & 0      & 0      & M_I^B  & M_I^B  & \cdots & M_I^B  & M_I^B  & M_I^B  & M_I^B  & \cdots & M_I^B  & M_S^B  
\end{vmatrix}.
\end{equation}

We again note the high degree of symmetry, which allows us reduce the determinant to
\begin{multline}
	0 = \left(\mathrm{det}\left|M_S^A-M_I^A\right|\right)^{N_a-1} \left(\mathrm{det}\left|M_S^B-M_I^B\right| \right)^{N_b-1} \left(\mathrm{det}\left| M_S^C - M_I^C \right| \right)^{N_c-1} \\ \times \mathrm{det}\left|M_S^A + (N_a-1)M_I^A\right| \mathrm{det}\left| M_S^B + (N_b-1)M_I^B\right| \\ \times \mathrm{det}\left| M_S^C + (N_c-1)M_I^C - N_c Z \right|,
\end{multline}
where
\begin{equation}
        Z = N_a M_I^A \left( M_S^A + (N_a-1) M_I^A \right)^{-1} M_I^A + N_b M_I^b \left( M_S^B + (N_b-1)M_I^B\right)^{-1} M_I^B.
\end{equation}
We recognize this as $N_a-1$ independent $A$ eigenvalues, $N_b-1$ independent $B$ eigenvalues, $N_c-1$ independent $C$ eigenvalues, one eigenvalue corresponding to the collective modes without the coupling to $C$ modes for each of the $A$ and $B$ modes, and a collective set for the $C$ modes with a modification due to the couplings to the $A$ and $B$ modes (through $Z$). We further note that when $N_c \rightarrow 0$, the eigenvalues reduce to two separate collective sets, as expected.

The interesting eigenvalue is due to the interaction between the $C$ modes' collective set and the couplings to $A$ and $B$ modes. 
If we assume that all mode parameters are the same for all sets of modes, and further assume that $N_a=N_b=N_c$, we can make analytic progress on this determinant, and obtain
\begin{equation}
	\left(\gamma + s\right)^2 + \Delta^2 - \omega^2 A_0^2 \left( (k_s + 2(N-1)k)^2 + 8 N^2 k^2 \right) = 0
\end{equation}
and the threshold amplitude
\begin{subequations}
	\begin{align}
		A_\mathrm{thr}^2 & = \frac{ \gamma^2 + \Delta^2 }{4 \omega^2 \left( 3k^2 N^2 + k(k_s-2k)N + k(k-k_s) + k_s^2/4\right)} \\
		               & \approx \frac{1}{3N^2} \left(\frac{ \gamma^2 + \Delta^2 }{ 4\omega^2 k^2 }\right) = \frac{1}{N_a^2 + N_b^2 + N_c^2} \left(\frac{ \gamma^2 + \Delta^2 }{ 4\omega^2 k^2 }\right),
	\end{align}
\end{subequations}
where we assumed the limit of large $N$. We note that this is very similar to the case of a single collective set, except $N^2 \rightarrow N_a^2 + N_b^2 + N_c^2$.
If we stitch together many separate collective sets by overlapping them, we only expect the effective number of modes to sum in quadrature.
This was tested numerically by taking the determinant without assuming equal numbers of modes, and found to be in reasonable agreement with this scaling.

\subsection{Non-``self coupled'' collective sets}\label{s:non-self coupled collective sets}

Appendix \ref{s:single collective set} and \ref{s:overlapping collective set} considered self-coupled modes.
 However, the vast majority of couplings will be between modes that do not support self-coupled daughters.
For example, if the parent azimuthal order $m$ is odd, then the daughter modes must have different $m$ numbers.
 If we consider two sets of modes, one with $N$ daughters and one with $n$ daughters, we can define $2\times2$ sub-matrices similar to Appendix \ref{s:single collective set} for each group of modes. 
 This means we will also find collective sets with characteristic equations like the following, with capital letters corresponding to the $N$-mode set and lower case letters corresponding to the $n$-mode set
\begin{equation}
0 = \mathrm{det}
\begin{vmatrix}
M_S    & 0      & \cdots & 0      & 0      & M_I    & M_I    & \cdots & M_I    & M_I    \\
0      & M_S    & \cdots & 0      & 0      & M_I    & M_I    & \cdots & M_I    & M_I    \\
\vdots & \vdots &        & \vdots & \vdots & \vdots & \vdots &        & \vdots & \vdots \\
0      & 0      & \cdots & M_S    & 0      & M_I    & M_I    & \cdots & M_I    & M_I    \\
0      & 0      & \cdots & 0      & M_S    & M_I    & M_I    & \cdots & M_I    & M_I    \\
M_i    & M_i    & \cdots & M_i    & M_i    & M_s    & 0      & \cdots & 0      & 0      \\
M_i    & M_i    & \cdots & M_i    & M_i    & 0      & M_s    & \cdots & 0      & 0      \\
\vdots & \vdots &        & \vdots & \vdots & \vdots & \vdots &        & \vdots & \vdots \\
M_i    & M_i    & \cdots & M_i    & M_i    & 0      & 0      & \cdots & M_s    & 0      \\
M_i    & M_i    & \cdots & M_i    & M_i    & 0      & 0      & \cdots & 0      & M_s    
\end{vmatrix}
\end{equation}
where this is an $(N+n)\times(N+n)$ matrix. We can simplify this to only a $4\times4$ determinant
\begin{equation}
0 = \left(\mathrm{det}\left|M_s\right|\right)^{N-1} \left(\mathrm{det}\left|M_s\right|\right)^{n-1} \mathrm{det}\begin{vmatrix} M_S & n M_I \\ N M_i & M_s \end{vmatrix},
\end{equation}
which looks like a set of independent eigenmodes and a $4\times4$ determinant for the collective modes.
In general, that $4\times4$ determinant must be solved numerically. However, if we again assume identical mode parameters and that $N=n$, we see that this reduces to 
\begin{equation}
\mathrm{det}\begin{vmatrix} M_S & N M_I \\ N M_I & M_S \end{vmatrix},
\end{equation}
which looks just like the three-mode instability equations with $k\rightarrow Nk$. 
Therefore, we can read off the amplitude threshold immediately.
Again, we see that the threshold is decreased by a factor of $N$ compared to the three-mode threshold.
We expect the threshold energy to approximately scale as
\begin{equation}
	A_\mathrm{thr}^2 \approx \left( \frac{1}{Nn} \right) \frac{ \gamma^2 + \Delta^2}{4\omega^2 k^2}.
\end{equation}

\subsection{Decoupling of ``very different'' modes from collective sets}\label{s:decoupling of large detuning}

In general, since all the mode parameters will be slightly different, our previous examples are a bit artificial.
We now investigate the behavior when one mode begins to differ from the others.
Consider the following characteristic equation, with $N$ identical modes and one slightly different mode indicated by $\delta M$
\begin{equation}
0 = \mathrm{det}
\begin{vmatrix}
M_S & M_I & \cdots & M_I & M_I \\
M_I & M_S & \cdots & M_I & M_I \\
\vdots & \vdots & & \vdots & \vdots \\
M_I & M_I & \cdots & M_S & M_I \\
M_I & M_I & \cdots & M_I & M_S + \delta M
\end{vmatrix}
\end{equation}
We can reduce this to 
\begin{subequations}
	\begin{align}
		0 & = \left(\mathrm{det}\left| M_S - M_I\right|\right)^{N-1} \mathrm{det} 
		\begin{vmatrix}
		M_S - M_I & - \delta M \\
		N M_I & M_S + \delta M
		\end{vmatrix} \\
		& = \left(\mathrm{det}\left| M_S - M_I\right|\right)^{N-1} \mathrm{det}\left|M_S + (N-1)M_I\right| \mathrm{det}\left| M_S - N M_I \left(M_S+(N-1)M_I\right)^{-1}M_I + \delta M \right|.
	\end{align}
\end{subequations}
\noindent
As $\delta M\rightarrow 0$, this reduces to a single collective set with $N\rightarrow N+1$, as expected. 
We also note that this looks like the eigenvalues of a normal collective set with $N$ modes and a new eigenvalue related to the different mode.
Furthermore, if $\delta M$ dominates the new eigenvalue, then we see that the different mode will ``decouple'' from the other modes.
Clearly, there will be some threshold for how large $\delta M$ needs to be before the different mode decouples, and that threshold will depend on the parent's amplitude in a non-trivial way.  
We expect that a large parent amplitude $A_o$ will support a larger $\delta M$ before the mode decouples.

\section{Scaling of parametric instability threshold and wave breaking threshold}\label{s:breaking thr}

Our calculations treat the system of modes as a set of global standing waves.  
However, if a wave's nonlinearity parameter $k_r \xi_r \ga 1$, the wave will invert the stratification of the star and break \citep{Goodman:1998, Barker:2010}.  
Because it does not reflect at turning points within the propagation cavity, such a wave is more appropriately treated as a traveling wave rather than a standing wave. 
Given that we specifically focus on parent waves below the wave breaking threshold ($k_r\xi_r\la 1$), we know that the parent is well described as a standing wave. 
Here we are interested in determining whether the same is true of the daughters, granddaughters, etc.  

As we describe in \S~\ref{s:selecting three-mode triples}, the parametric instability threshold scales as $E_{\rm thr} \propto \omega^6$. 
This implies that each successive generation has a lower $E_{\rm thr}$ and is therefore ever more susceptible to parametric instabilities. 
We show below that the energy above which a wave breaks also scales as $E_{\rm break}\propto \omega^6$. 
Moreover, we find that $E_{\rm thr} \ll E_{\rm break}$. 
This means that well before the daughters, granddaughters, etc. reach the wave breaking limit $k_r\xi_r \ga 1$, they will excite the next generation of modes through parametric instabilities.   
Although a mode is not necessarily limited to remain below its $E_{\rm thr}$, we do not expect it to greatly exceed it either.  
This is because as a mode's amplitude increases past its $E_{\rm thr}$, its children grow at an ever faster rate and thereby limit how far their parent overshoots $E_{\rm thr}$.
While this issue requires further study, it suggests that our assumption that the modes are all global standing waves may be reasonable.

We begin by calculating $E_{\rm thr}$.  For typical parameter values of a hot Jupiter system, $E_{\rm thr}$ is limited by the nonlinear detuning of the daughter modes rather than their linear damping (and similarly for granddaughters, etc.).  
To a first approximation, the detuning $\Delta$ is determined by half the frequency spacing between the daughter modes $\omega / 2n$. 
However, this assumes that the lowest $E_{\rm thr}$ pairs are self-coupled modes.  
Because there is a distribution of mode frequencies slightly above and below half the parent frequency, there are always some mode pairs that happen to have $\Delta \ll \omega/2n$  \citep{Wu:2001}.  
These are the pairs that minimize $E_{\rm thr}$.  
Writing $\Delta =\alpha \omega/2n$, where $\alpha \ll 1$ and using the expressions for $\omega$, $\gamma$, and $\kappa$ given in \S~\ref{s:properties of the modes}, we find that the threshold energy for self-coupled daughters is
\begin{equation}
E_{\rm thr} \simeq 8\times10^{-16} \left(\frac{\alpha}{0.01}\right)^2\left(\frac{P}{\trm{day}}\right)^{-6} E_0,
\end{equation}
where $\alpha\sim 0.01$ based on our three-mode network search results (cf. Figure \ref{f:Athr distribution}).

Now consider $k_r \xi_r$. It is at its maximum near the inner turning point of the parent (where $\omega \simeq N$).  
This is because in the core of a solar model, $k_r\simeq \Lambda N/\omega r$ is approximately constant and $\xi_r \propto r^{-2}$ by flux conservation.  
Using the WKB relations given in Appendix A of WAQB  (see also \citealt{Goodman:1998, Ogilvie:2007}), we find that the wave breaking condition $\max \{k_r \xi_r\}=1$ for $l=2$ modes corresponds to an energy 
\begin{equation}
E_\mathrm{brk} \simeq 3 \times 10^{-13} \left(\frac{P}{\trm{day}}\right)^{-6} E_0.
\end{equation}
\noindent
Longer period modes break at lower amplitudes because they reach further into the core of the star.  
We thus see that both energies scale as $\omega^6$ and $E_{\rm thr} \ll E_{\rm brk}$, as claimed.

\section{estimate of the linear and parent-daughter orbital decay timescales}\label{s:Tlin}

 The linear dissipation rate of individual resonant modes is $\dot{E}_{\rm lin}\simeq 2\gamma_\alpha E_{\rm lin}$, where $E_{\rm lin}$ is given by Equation (\ref{e:Elin}). For the short periods that we consider, $\Delta_\alpha \approx \omega_\alpha/2n_\alpha \gg \gamma_\alpha$. After summing over many parents near the resonance, 
 using the WKB estimates for the damping and forcing coefficients (Equations \ref{e:gamma_approx} and \ref{e:forcing}), and averaging according to Appendix \ref{s:time averaging}, we find
\begin{equation}
	\aveT_{\rm lin} \simeq 1.4\times10^{12} \left(\frac{M}{M_\odot}\right)^{-5/6}  \left(\frac{R}{R_\odot}\right)^{11/2} \left(\frac{\Mp}{\Mj}\right)^{-1}\left(\frac{\Porb}{\rm day}\right)^3\trm{ yr}
\end{equation}
\noindent
and 
\begin{equation}
	Q^\prime_{\ast,\rm lin} \simeq 1.1 \times 10^{10} \left(\frac{R}{R_\odot}\right)^{21/2} \left(\frac{M}{M_\odot}\right)^{-27/6} \left(\frac{\Mp}{\Mj}\right) \left(\frac{\Porb}{\mathrm{day}}\right)^{-4/3}.
\end{equation}

As we explain in \S~\ref{s:parents and daughters}, we can also estimate the nonlinear dissipation rate of networks consisting of only parents and daughters (but not granddaughters, etc.).  This is because the dissipation in that case is dominated by the single daughter pair $(\beta,\gamma)$ with the lowest instability threshold $E_{\rm thr}$.  As we show in Appendix \ref{s:3mode equilib derivation}, for the parameters of a hot Jupiter system, the nonlinear equilibrium energy of such a daughter pair is $E_{\beta,\gamma} \simeq |U_\alpha/2\kappa_{\alpha \beta \gamma}|E_0$. The total dissipation rate of the system is approximately the dissipation due to these two daughters $\dot{E}_{\textrm{p-d}}\simeq 2\times 2\gamma_{\beta,\gamma} E_{\beta,\gamma}$. 
There is a small correction to this because the lowest $E_{\rm thr}$ daughters have slightly different parameters and therefore do not sit at exactly the same amplitudes.
After accounting for this small correction and plugging in Equations \ref{e:gamma_approx}, \ref{e:forcing} and \ref{e:kappa_approx}, we find
\begin{equation}
	\aveT_{\textrm{p-d}} \simeq 2.0\times10^{11} \left(\frac{\Lambda_{\beta,\gamma}^2}{2}\right)^{-1} \left(\frac{T_{\alpha\beta\gamma}}{0.2}\right) \left(\frac{M}{M_\odot}\right)^{-11/6} \left(\frac{R}{R_\odot}\right)^{11/2} \left(\frac{\Porb}{\rm day}\right)^{19/6}\trm{ yr}
\end{equation}
and
\begin{equation}
	Q^\prime_{\ast,\textrm{p-d}} \simeq 1.5 \times10^{9}\left(\frac{\Lambda_{\beta,\gamma}^2}{2}\right)^{-1} \left(\frac{T_{\alpha\beta\gamma}}{0.2}\right)\left(\frac{M}{M_\odot}\right)^{-27/6} \left(\frac{R}{R_\odot}\right)^{21/2} \left(\frac{\Mp}{\Mj}\right)  \left(\frac{\Porb}{\rm day}\right)^{-7/6}. 
\end{equation}
Here we took $l_{\beta,\gamma}=1$, which is representative of the typical lowest $E_{\rm thr}$ daughters for $P\ga 2\trm{ days}$.
We find good agreement between the parent-daughter network integrations that include many daughters and this analytic estimate (see circles and dashed curve in Figure \ref{f:frequency sweeps}).  In the figure, we assume $l_{\beta,\gamma}=1$ even for $P<2\trm{ days}$.  However, at these shorter periods, the available daughter modes are spaced further apart in frequency and the lowest $E_{\rm thr}$ pair may be pushed to  $l_{\beta, \gamma} \ga 1$.  This causes the small discrepancy between the circles and dashed curve at $P\la 2\trm{ days}$ seen in Figure \ref{f:frequency sweeps}.

\bibliography{refs}

\end{document}